\documentclass[useAMS,usenatbib,usegraphicx]{mn2e}

\usepackage{subfigure}
\usepackage{url}
\usepackage{comment}
\usepackage{dcolumn}
\usepackage{enumitem}
\usepackage{hyperref}
\usepackage{longtable}
\usepackage{color}
\usepackage[table]{xcolor}
\usepackage{colortbl}
\usepackage{amsmath}

\newcommand*{\rom}[1]{\uppercase\expandafter{\romannumeral #1}}
\newcommand\ion[2]{#1$\;${\small\rmfamily\rom{#2}}\relax}%

\newcommand{\myAA}{\AA \hspace{0.025in}}
\newcommand{\new}[1]{{\color{black}{\;#1\;}}}

\newcommand\aap{{A\&A}}%
\newcommand\mnras{{MNRAS}}%
\newcommand\apj{{ApJ}}%
\newcommand\apjs{{ApJS}}%
\newcommand\apjl{{ApJ}}%
\newcommand\aj{{AJ}}%
\newcommand\pasp{{PASP}}%
\newcommand\nat{{\it Nature}}%

\setlist[enumerate]{noitemsep}
\setlist[enumerate,1]{leftmargin=*}
\setlist[itemize]{noitemsep}
\setlist[itemize,1]{leftmargin=*}
\setlist[description]{noitemsep}
\setlist[description,1]{leftmargin=*}
\setenumerate[0]{label=(\arabic*)}

\makeatother

\title[M31 GC Metallicities]{Metallicities of Outer Halo M31 Globular Clusters from Integrated Light Calcium-II Triplet Spectroscopy}

\author[Sakari et al.]{Charli M. Sakari$^{1}$\thanks{E-mail:
    sakaricm@sfsu.edu} and George Wallerstein$^{2}$\\
$^{1}$ San Francisco State University, 1600 Holloway Avenue, San Francisco, CA, 94132 USA\\
$^{2}$ Department of Astronomy, University of Washington, Seattle WA
98195-1580, USA\\
}

\begin{document}

\maketitle

\label{firstpage}

\begin{abstract}
This paper presents [Fe/H] ratios for GCs in the outer halo of the Andromeda Galaxy, M31, based on moderate-resolution, integrated light (IL) spectroscopy of the calcium-II triplet (CaT) lines.  The CaT strengths are measured by fitting Voigt profiles to the lines and integrating those profiles; integrations of defined bandpasses are also considered.  The [Fe/H] ratios are determined using an empirical calibration with CaT line strength, as derived from another sample of M31 GCs that were previously studied at high-resolution.  The [Fe/H] ratios for the new GCs reveal that the outer halo GCs are indeed generally more metal-poor than typical inner halo GCs, though there are several more metal-rich GCs that look to have been accreted from dwarf satellites.  The metallicities of these GCs also place important constraints on the nature of the substructure in the outer halo and the dwarf satellites that created this substructure.
\end{abstract}


\section{Introduction}\label{sec:Intro}
Observational knowledge of galaxy assembly has expanded dramatically over the last few decades.  Large-scale surveys (such as the Sloan Digital Sky Survey; \citealt{SDSSRef}) have examined large numbers of galaxies, while deep surveys of the Milky Way (e.g., {\it Gaia}; \citealt{GaiaREF}) have provided detailed knowledge about a single galaxy.  However, much work remains to investigate the detailed properties of galaxies beyond the Milky Way and its satellites and, ultimately, to place the Milky Way in a cosmological context.  As the nearest large spiral galaxy to the Milky Way, M31 is an ideal laboratory for such detailed studies, particularly in comparison with the Milky Way.

M31 has been observed since antiquity, but became a particularly useful case for galaxy studies at the beginning of the twentieth century \citep{Hubble1929}.  Modern understanding of M31 has benefited greatly from large photometric and spectroscopic surveys, including the Panchromatic Hubble Andromeda Treasury (PHAT; \citealt{Dalcanton2012}), the Spectroscopic and Photometric Landscape of Andromeda’s Stellar Halo (SPLASH; e.g., \citealt{Kalirai2006}), and the Pan-Andromeda Archaeological Survey (PAndAS; \citealt{McConnachie2009}).  Among other things, these surveys have revealed complex stellar populations within M31 that are unlike those in the Milky Way, particularly in the outer halo, which suggests it has had an active and recent merger/accretion history. In the outer halo, which is defined as the region with a projected distance from the centre greater than 25 kpc,  coherent stellar streams and intact satellite galaxies are plentiful, revealing the complexity of M31's ongoing assembly \citep{McConnachie2018}.  Detailed studies of the nature of the accreted satellite galaxies (including their masses, stellar compositions, and ages) will be essential for comparisons with the Milky Way and galaxy formation simulations.

The globular clusters (GCs) in M31 have provided unique and valuable information about the properties of these accreted satellites \citep{Mackey2019} as well as {\it in situ} cluster formation within M31 itself (e.g., \citealt{Caldwell2009,Caldwell2011,Schiavon2012,Schiavon2013,Johnson2017}).  Although many open questions remain about GC formation, GCs have been shown to be ubiquitous in sufficiently massive galaxies (e.g., \citealt{Harris2013}).  GCs trace the age-metallicity relations and chemical properties of their birth environments (e.g., \citealt{Pritzl2005,Letarte2006,Hendricks2016}) and are believed to trace a galaxy's assembly history (e.g., see the simulations from \citealt{Kruijssen2019}, \new{\citealt{Hughes2019}}).  GCs are also observable at greater distances than individual stars and can be detected when a background stream is too faint or diffuse.  The GCs in M31 are close enough to be partially resolved for photometry (e.g., \citealt{Mackey2013}) and low-resolution spectroscopy (e.g., \citealt{Reitzel2004}), but the brightest red giant branch (RGB) stars are too faint for chemical abundance studies at high spectral resolution, at least with current telescopes and instrumentation.  Instead, GCs can be studied with Integrated Light (IL) spectroscopy, which yields radial velocities, metallicities, and ages (at lower resolution; e.g., \citealt{Caldwell2009,Schiavon2012,Schiavon2013}) as well as detailed chemical abundances (at higher resolution; e.g., \citealt{Colucci2009,Colucci2014,Sakari2013,Sakari2015,Sakari2016}, \new{\citealt{Larsen2018,Larsen2021,Larsen2021b}}).

M31's GCs have already revealed a wealth of information about the galaxy's assembly, particularly in the outer halo.  \new{Simulations suggest that GCs with orbits in the outer halo are likely to have been accreted either fairly recently or from old, low mass systems \citep{Pfeffer2020}.  \citet{Hughes2019} also show that younger, metal-rich GCs are likely to have been accreted from higher-mass dwarf satellites.  The observations of M31's outer halo do show strong evidence for ongoing accretion of dwarf galaxies and their GCs.} \citet{Mackey2019} note that many GCs show strong associations with bright stellar streams, based on positions or velocities (also see \citealt{Veljanoski2014}).  The GCs that are not associated with substructure appear to be from older accretion events \citep{Mackey2019b}, while the others are likely from more recent accretion events \citep{McConnachie2018}.  Several of the outer halo GCs have been spectroscopically studied, providing metallicities or detailed chemical abundances \citep{Colucci2009,Colucci2014,Sakari2015,Sakari2016,Sakari2021,SakWall2016,Larsen2018,Larsen2021,Larsen2021b}, but many more remain to be studied.

This paper presents metallicities for thirty GCs in the outer halo of M31, based on the IL strengths of the near-infrared calcium-II triplet (CaT) lines.  The CaT lines in IL spectra are known to trace the overall metallicity of a cluster (e.g., \citealt{AZ88,SakWall2016,Usher2019}).  Unlike IL spectra that go further into the blue, the CaT lines are much less sensitive to the GC age \citep{SakWall2016,Usher2019}, avoiding the well-known age-metallicity degeneracy that affects IL spectra (see, e.g., \citealt{Schiavon2004}).  This paper presents the first CaT-based metallicities for these GCs.

Section \ref{sec:Observations} describes the observations and data reduction, while Section \ref{sec:AnalTechniques} outlines the techniques for measuring the strengths of the CaT lines.  The metallicities are presented in Section \ref{sec:FeH}, along with considerations of whether to use [Fe/H] or [Ca/H] and comparisons with values from the literature.  In Section \ref{sec:Discussion} the metallicities of the outer halo GC system as a whole are discussed, along with the properties of individual streams and groups of GCs.  The final conclusions are then summarized in Section \ref{sec:Conclusion}.

\section{Observations and Data Reduction}\label{sec:Observations}
Spectra of thirty outer-halo GCs in M31 were obtained with the Astrophysical Research Consortium 3.5-m telescope at Apache Point Observatory in 2015, 2019, and 2020.  Clusters were prioritized based on total brightness, IL colour, projected distance from the centre of M31, and potential associations with specific streams.  These targets and their magnitudes are listed in Table \ref{table:Targets}.

\begin{table*}
\centering
\begin{center}
\caption{Observed Targets.\label{table:Targets}}
\newcolumntype{d}[1]{D{,}{\;\pm\;}{#1}}
\begin{tabular}{@{}lccccccd{3}c@{}}
  \hline
Cluster & RA$^{a}$ & Dec & $M_V$ & Date of & Exposure & S/N$^{b}$ & \multicolumn{1}{c}{$v_{\rm{helio}}$} & Notes$^{c}$ \\
   & (J2000) & (J2000) & & Observation & Time (s) & (8525 \AA) & \multicolumn{1}{c}{(km/s)} & \\
  \hline
PA-01 & 23:57:12.0 & $+$43:33:08.3 & -7.48 & 2015 Nov 11,$^{d}$ Dec 11$^{d}$ & 7200 & 30 & -389.7,17.5 & \\
PA-04 & 00:04:42.9 & $+$47:21:42.5 & -7.09 & 2015 Aug 13, 20                 & 8400 & 44 & -396.3,12.3 & NW Stream \\
PA-09 & 00:12:54.7 & $+$45:05:55.9 & -6.75 & 2019 Nov 3                      & 4320 & 50 & -442.0,10.9 & NW Stream \\
PA-11 & 00:14:55.6 & $+$44:37:16.4 & -6.74 & 2019 Oct 6                      & 4800 & 16 & -405.7,20.2 & NW Stream \\
PA-14 & 00:20:33.9 & $+$36:39:34.5 & -7.01 & 2015 Nov 11$^{d}$               & 4800 & 22 & -372.6,15.0 & SW Cloud \\
PA-16 & 00:24:59.9 & $+$39:42:13.1 & -8.44 & 2015 Aug 13                     & 3600 & 74 & -469.5,6.4 & \\
H4    & 00:29:45.0 & $+$41:13:09.4 & -7.82 & 2019 Nov 5, 2020 Sep 27         & 4500 & 80 & -382.3,7.7 & \\
H5    & 00:30:27.3 & $+$41:36:19.5 & -8.44 & 2019 Oct 6                      & 3600 & 44 & -397.4,10.7 & \\
H7    & 00:31:54.6 & $+$40:06:47.8 & -7.17 & 2019 Nov 3                      & 4329 & 36 & -425.0,15.7 & Association 2\\
PA-22 & 00:32:08.4 & $+$40:37:31.6 & -6.18 & 2019 Nov 3                      & 6000 & 26 & -440.5,17.5 & Association 2\\
PA-27 & 00:35:13.5 & $+$45:10:37.9 & -7.69 & 2015 Aug 20                     & 3600 & 44 & -29.6,13.2 & \\
dTZZ-05&00:36:08.6 & $+$39:17:30.0 & -7.03 & 2019 Aug 01                     & 5700 & 64 & -61.5,7.6 & Association 2?\\
H11   & 00:37:28.0 & $+$44:11:26.5 & -7.88 & 2019 Nov 3                      & 2700 & 66 & -222.5,5.3 & \\
H12   & 00:38:03.9 & $+$37:44:00.2 & -8.19 & 2019 Oct 6                      & 3600 & 122 & -387.0,5.3 & \\
H18   & 00:43:36.1 & $+$44:58:59.3 & -8.09 & 2020 Sep 27                     & 1800 & 60 & -237.7,15.7 & \\
H19   & 00:44:14.9 & $+$38:25:42.2 & -7.29 & 2019 Oct 6                      & 4800 & 40 & -272.8,13.7 & GSS? \\
PA-36 & 00:44:45.6 & $+$43:26:34.8 & -7.30 & 2015 Nov 11$^{d}$               & 4800 & 38 & -417.9,15.5 & Substructure?\\
G339  & 00:47:50.2 & $+$43:09:16.5 & -7.58 & 2020 Sep 27                     & 2700 & 44 & -84.6,10.2 & Substructure?\\
PA-37 & 00:48:26.5 & $+$37:55:42.1 & -7.35 & 2015 Nov 11$^{d}$               & 4800 & 52 & -393.4,10.3 & GSS\\
H22   & 00:49:44.7 & $+$38:18:37.8 & -7.65 & 2019 Sep 28                     & 3600 & 64 & -302.0,9.4 & GSS? \\
EXT8  & 00:53:14.5 & $+$41:33:24.5 & -9.28 & 2019 Sep 28                     & 2577 & 108 & -208.2,10.2 & \\
PA-41 & 00:53:39.6 & $+$42:35:15.0 & -7.07 & 2019 Nov 3                      & 3600 & 24 & -142.8,16.2 & Stream C/D\\
H24   & 00:55:43.9 & $+$42:46:15.9 & -7.10 & 2019 Nov 5                      & 5700 & 34 & -129.0,16.5 & Stream C/D\\
PA-44 & 00:57:55.9 & $+$41:42:57.0 & -7.72 & 2015 Aug 20, 2015 Nov 11$^{d}$  & 5100 & 94 & -359.7,13.0 & Stream C/D\\
PA-46 & 00:58:56.4 & $+$42:27:38.3 & -8.67 & 2015 Aug 20, 2020 Sep 27        & 3900 & 84 & -117.9,19.4 & Stream C/D\\
H25   & 00:59:34.6 & $+$44:05:38.9 & -7.93 & 2019 Nov 5, 2020 Sep 27         & 3300 & 64 & -203.7,5.2 & \\
B517  & 00:59:59.9 & $+$41:54:06.8 & -8.17 & 2019 Nov 3                      & 2700 & 66 & -267.5,8.3 & Stream C/D\\
H27   & 01:07:26.3 & $+$35:46:48.4 & -8.39 & 2019 Nov 5                      & 2700 & 88 & -279.6,11.6 & \\
PA-52 & 01:12:47.0 & $+$42:25:24.9 & -7.58 & 2015 Oct 17                     & 3370 & 30 & -294.0,16.9 & \\
dTZZ-21&01:28:49.2 & $+$47:04:22.0 & -7.25 & 2019 Nov 5, 2020 Sep 27         & 7200 & 44 & -265.8,11.8 & \\
\hline
\end{tabular}
\end{center}
\medskip
\raggedright
$^{a}$ Positions and total V band magnitudes are from \citet{Mackey2019}.\\
$^{b}$ S/N ratios are per resolution element.\\
$^{c}$ Potential stream associations are from \citet{Mackey2019}.  Question marks indicate an ambiguous association with substructure.\\
$^{d}$ These nights had issues with the wavelength solutions (see the text).\\
\end{table*}

The Dual-Imaging Spectrograph (DIS) with the R1200 grating and a slit width of 1.\arcsec5 was used to obtain moderate-resolution spectra with a spectral resolution of 0.56 \AA/pix (or $R\sim 4,000$ at 8500 \AA), covering $\sim8000 - 9100$ \AA $\;$ on the red camera.  The data from the blue camera was not utilized.  Exposure times per visit were 15-20 minutes; the aim was to obtain S/N ratios of $\sim20$ per resolution element for each cluster, though this goal was not always achieved.  The final exposure times and S/N ratios are shown in Table \ref{table:Targets}.

\begin{figure}
\begin{center}
\centering
\hspace*{-0.25in}
\includegraphics[scale=0.58,trim=0.0in 0.0in 0.5in 0.4in,clip]{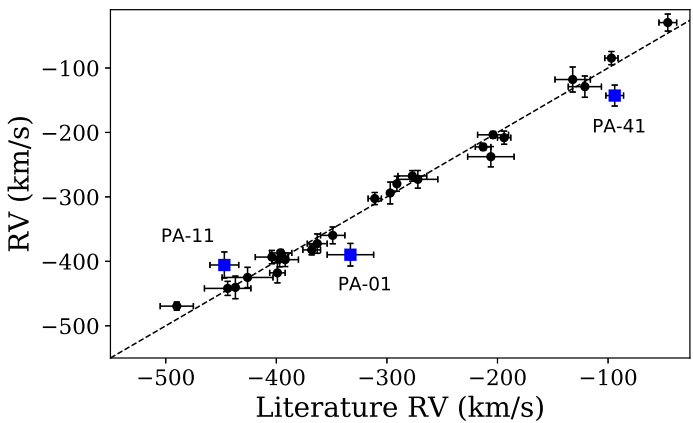}
\caption{Comparisons between heliocentric radial velocities from this paper ($y$-axis) and those from \citet[$x$-axis]{Veljanoski2014}.  Four clusters are found to not agree within 1$\sigma$ uncertainties; they are shown with blue squares and are labelled.}\label{fig:RVs}
\end{center}
\end{figure}

The data were reduced in the Image Reduction and Analysis Facility program (IRAF)\footnote{IRAF is distributed by the National Optical Astronomy Observatory, which is operated by the Association of Universities for Research in Astronomy, Inc., under cooperative agreement with the National Science Foundation.} using standard techniques, including variance weighting, as described in \citet{SakWall2016}.  Sky subtraction is crucial for the CaT lines, particularly the third CaT line, which lies in a cluster of sky emission lines.  The sky spectra were selected from regions on either side of the cluster; the IRAF task \textit{skytweak} was used to align the sky emission lines with the object spectra to optimize the sky subtraction.  Radial velocities were then determined through cross-correlations with a high resolution, high S/N spectrum of Arcturus \citep{Hinkle2003}\footnote{\url{ftp://ftp.noao.edu/catalogs/arcturusatlas/}} that had been downgraded to the resolution of DIS.  The final heliocentric radial velocities for all GCs are shown in Table \ref{table:Targets}.  Two nights in 2015 were found to have large radial velocity offsets in the sky lines, indicating issues with their wavelength solutions; for these spectra (which are identified in Table \ref{table:Targets}) the night sky lines were used to determine the approximate velocity offsets needed to correct the M31 GC velocities.  The errors in the velocities for all GCs are calculated by \new{adding in quadrature the radial velocity uncertainty from the cross-correlation,  the standard deviation between individual exposures, and the uncertainties from the wavelength solutions}.  Figure \ref{fig:RVs} shows comparisons with the literature values from \citet{Veljanoski2014}.  Three GCs have slightly discrepant RVs from the \citet{Veljanoski2014} values, as seen in Figure \ref{fig:RVs}.  \new{Two (PA-11 and PA-01) agree with the literature within $2\sigma$, while PA-41 agrees within $3\sigma$.}  Only one of these discrepant clusters, PA-01, was observed in late 2015, on a night with a poor wavelength calibration, suggesting that the wavelength solution issues cannot be solely responsible for the offsets.  \new{It is likely that random effects are not being completely accounted for in the calculated uncertainties, particularly for PA-41.}  Note that dTZZ-05 and dTZZ-21 were discovered more recently \citep{dTZZ} and this paper presents the first RVs for these clusters.\footnote{Note that it is possible that a foreground Milky Way star was observed instead of dTZZ-05; see Section \ref{subsubsec:dTZZ-05}.}

\begin{figure}
\begin{center}
\centering
\hspace*{-0.15in}
\includegraphics[scale=0.45,trim=0in 0.5in 1.0in 0.75in,clip]{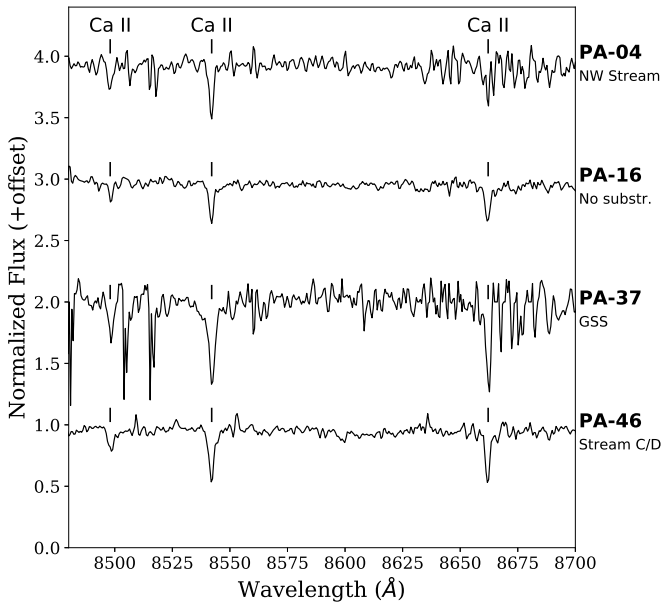}
\caption{Sample CaT spectra of four M31 outer halo GCs.  Possible stream associations are also identified.}\label{fig:ExampleSpectra}
\end{center}
\end{figure}

The spectra were normalized with very low-order polynomial fits and individual rest-frame exposures were combined with average sigma-clipping.  Examples of final spectra are shown in Figure \ref{fig:ExampleSpectra}.  The three CaT lines are labelled; the third CaT line is occasionally undetectable, as a result of the difficult sky subtraction.

\section{Measurements of CaT Line Strengths}\label{sec:AnalTechniques}
\citet{SakWall2016} present a lengthy discussion and comparison of different ways of measuring the strengths of the CaT lines, ultimately providing a summary for which measurements best track the high-resolution [Fe/H] ratios.  To briefly summarize, \citet{SakWall2016} identified two primary ways of measuring the CaT line strengths: 1) integrations of given bandpasses, or indices (also known as pseudo-equivalent widths, because the lower-resolution features are often blends) and 2) full Voigt profile fits.  Based on prior work with extragalactic CaT spectra (e.g., \citealt{Foster2011}, \citealt{Usher2012}), \citet{SakWall2016} also tested the technique of using template fits to raw spectra; this template fitting can mitigate problems with low S/N and improper sky subtraction, provided that the templates have been observed with the same setup.  Sakari \& Wallerstein found no significant systematic differences between measurements on raw vs. template-fitted spectra, regardless of whether integrations or Voigt profile fits were used.

This paper will build off the analysis of \citet{SakWall2016}, using their recommended measurement techniques (described in more detail below) on both raw and template-fitted spectra. Since the measurement methods are slightly modified from the original 2016 analysis, the 32 GCs from \citet{SakWall2016} and the 1 GC from \citet{Sakari2021} are also re-analyzed here, for consistency.

\subsection{Template Fits}\label{subsec:Templates}
\begin{figure}
\begin{center}
\centering
\hspace*{-0.15in}
\includegraphics[scale=0.55,trim=0.15in 0.0in 0.45in 0.4in,clip]{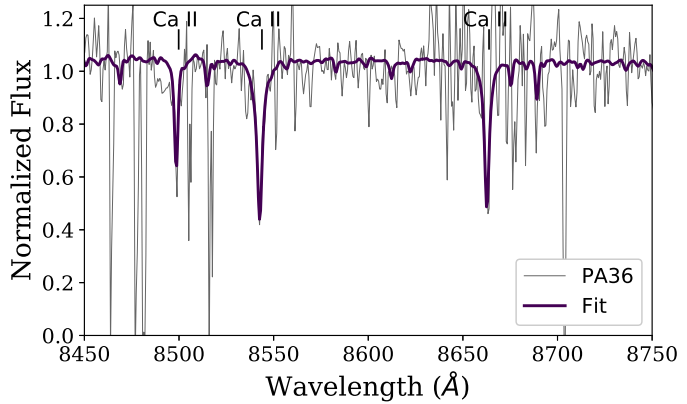}
\caption{An example template fit (thick purple line) to the cluster PA-36 (thin grey line).}\label{fig:Template}
\end{center}
\end{figure}

Template fits to the GC spectra were found with linear combinations of stellar spectra using the Penalized Pixel-Fitting \texttt{pPXF} code \citep{pPXFref}, as described in \citet{Foster2011} and \citet{SakWall2016}.  The template spectra are from bright Milky Way stars that span a range of temperatures and metallicities.  Figure \ref{fig:Template} shows an example of a template fit to the spectrum of PA-36; the fit to the first and second CaT lines is excellent, while the template vastly improves the detection of the third CaT line (which is obscured by poorly removed sky lines in the raw spectrum).

The line measurements described below are done on both the template-fitted and raw spectra, although the low S/N and sky line contamination in the raw spectra often lead to large uncertainties in the final CaT line strengths and, as a result, the [Fe/H] ratios.

\subsection{Pseudo EWs: Integrations of Bandpasses}\label{subsec:Integrations}
\begin{figure}
\begin{center}
\centering
\hspace*{-0.15in}
\includegraphics[scale=0.55,trim=0.15in 0.0in 0.5in 0.45in,clip]{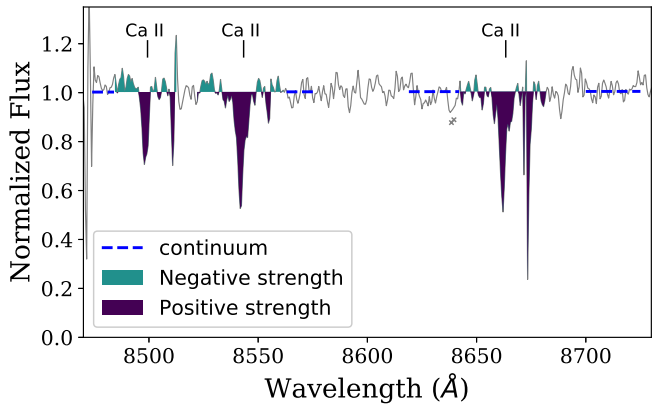}
\caption{An example of the area included in a pseudo-EW integration for the cluster B517.  The dashed blue lines show the C01 continuum regions and the resulting fit (note that the redder C01 continuum region is not shown); the crosses show points that were removed from the continuum fit.  The dark purple and cyan areas show the positive and negative areas that are included in the C01 index definitions for the three CaT lines. Both the first and third CaT lines show evidence of sky line contamination in the line bandpasses, which alter the resulting line strengths.}\label{fig:IntExample}
\end{center}
\end{figure}

The lines in these moderate-resolution spectra are generally blends of multiple spectral lines, making it difficult to measure EWs of the individual CaT lines through integrations.  However, one can obtain ``pseudo-EWs'' by integrating within a well-defined bandpass, or index.   \citet{SakWall2016} adopted the index definitions of \citet[hereafter C01]{Cenarro2001}, both for the continuum and line definitions, and these definitions are also adopted in this paper.  An example integration of the C01 bandpass is shown in Figure \ref{fig:IntExample}.  Although the spectra in this paper are normalized, the continuum definitions are essential for reliably and consistently measuring the line strengths.  These indices are typically measured on un-normalized spectra, but low-order normalization will not affect the strengths of spectral lines.  Note that an improperly removed sky line occasionally falls within the C01 continuum or index definitions, which can significantly alter the resulting line strength.  To solve this issue, significant outliers were removed from the initial continuum fits, as shown in Figure \ref{fig:IntExample}.

The integrations were done in Python using the trapezoidal method.\footnote{\url{www.scipy.org/}, \citet{SciPyREF}}  Errors in the line strengths were determined from the S/N of the spectrum and the resulting uncertainty in the area.  Note that this is slightly different than the procedure in \citet{SakWall2016}, who used a separate code (\texttt{indexf}; \citealt{Cardiel2010}) to measure index strengths.  The reliability of the new Python code was tested by comparing the C01 index measurements from \texttt{indexf} to the trapezoidal method in Python, using the 32 GCs from \citet{SakWall2016}.   The agreement between the two codes is generally quite good; any outliers are found to be a result of issues with continuum placement and residual sky lines.  The uncertainties in the line strengths are also quite similar between the two methods.  To account for the slight differences in the line measurements, the 32 GCs from \citet{SakWall2016} are re-analyzed here.  Any comparisons will utilize the measurements from this paper, rather than measurements from \citet{SakWall2016}.

GCs with low S/N and large residual sky lines have large uncertainties in their raw CaT strengths, particularly for the first and third CaT lines.  The first line is the weakest of the triplet and falls next to a strong sky emission line, while the third CaT line is located in a region with many strong sky lines.  The second CaT line, however, is generally relatively isolated, and can serve as a useful indicator of metallicity from the raw spectra (see Section \ref{subsec:Raw_vs_Temp}).

\subsection{Voigt Profile Fits}\label{subsec:Voigts}
\begin{figure}
\begin{center}
\centering
\hspace*{-0.15in}
\includegraphics[scale=0.55,trim=0.15in 0.0in 0.5in 0.4in,clip]{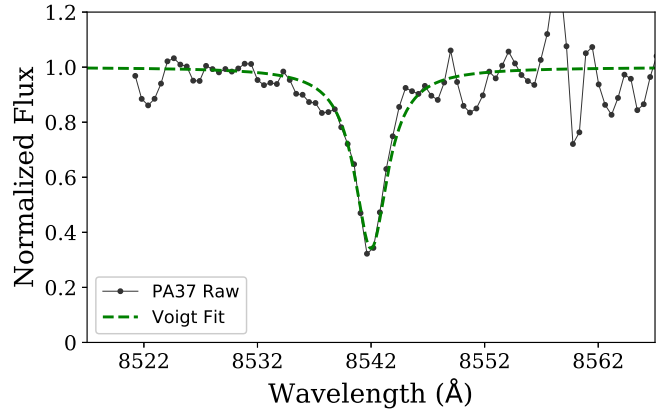}
\caption{An example Voigt profile fit to the second CaT line in the raw PA-37 spectrum.}\label{fig:PA37Voigt}
\end{center}
\end{figure}

As in \citet{SakWall2016}, spectral line strengths were also determined with Voigt profile fits, using the Python program {\tt pymodelfit}.\footnote{\url{https://pythonhosted.org/PyModelFit/}}  \citet{SakWall2016} showed that Voigt profiles better fit the CaT lines than Gaussian profiles, especially for the stronger second and third lines.  The Voigt fits were done with rough continuum fitting and masking of known nearby spectral features.  An example Voigt profile fit to the second CaT line in a raw spectrum is show in Figure \ref{fig:PA37Voigt}.  The strengths of the spectral lines were then determined by integrating the full Voigt profile.  Uncertainties in the resulting area of the Voigt profile were conducted with Monte Carlo resampling and bootstrapping of the spectra.  The spectra were resampled and refit 1000 times; the adopted error in the line strength is then the standard deviation of the measurements.

\section{Metallicities}\label{sec:FeH}
This section uses the line strength measurements from the previous section to re-derive a relationship between IL CaT strength and IL [Fe/H], using the re-measurements of the M31 GC sample from \citet{SakWall2016}; re-measurements are necessary because the measurement techniques are slightly different in this paper.  This relationship will then be used to determine metallicities for the new M31 GCs.   The empirical relationships were determined using linear fits to high-resolution [Fe/H] vs. CaT line strengths\footnote{Unlike \citet{SakWall2016}, this paper does not find a significant advantage to piecewise linear fits.} with the SciPy \citep{SciPyREF} optimize.curve\_fit package.\footnote{\url{http://www.scipy.org/}}  Only errors in one dimension can be included in the fit---here the errors in metallicity are adopted, since they have a greater effect than the uncertainties in the CaT strength (see Section \ref{subsec:CaH}).  The original GCs from \citet{SakWall2016} were used for this calibration; as in that previous analysis, B457 was not included in the calibration because of its discrepant radial velocity.

In order to identify the most robust way of measuring the IL CaT line strengths, Section \ref{subsec:Raw_vs_Temp} compares the derived relationships from the raw and the template-fitted spectra, while Section \ref{subsec:Voigt_vs_Int} compares the pseudo EWs to the Voigt profile fits.  Ultimately, the Voigt profile fits to the second CaT line in the template-fitted spectra are shown to provide the best fits to the calibration set of GCs.  Section \ref{subsec:CaH} then discusses whether [Fe/H] or [Ca/H] better trace the IL CaT strengths, using the previously-studied GCs from \citet{SakWall2016}.  The final adopted [Fe/H] ratios for the new M31 GCs are provided and are compared with literature values in Section \ref{subsec:FeHresults}.

\subsection{Raw vs. Template-Fitted Spectra}\label{subsec:Raw_vs_Temp}
\begin{figure*}
\begin{center}
\centering
\subfigure[Voigt Fits]{\includegraphics[scale=0.53,trim=0in 0in 0.45in 0.45in,clip]{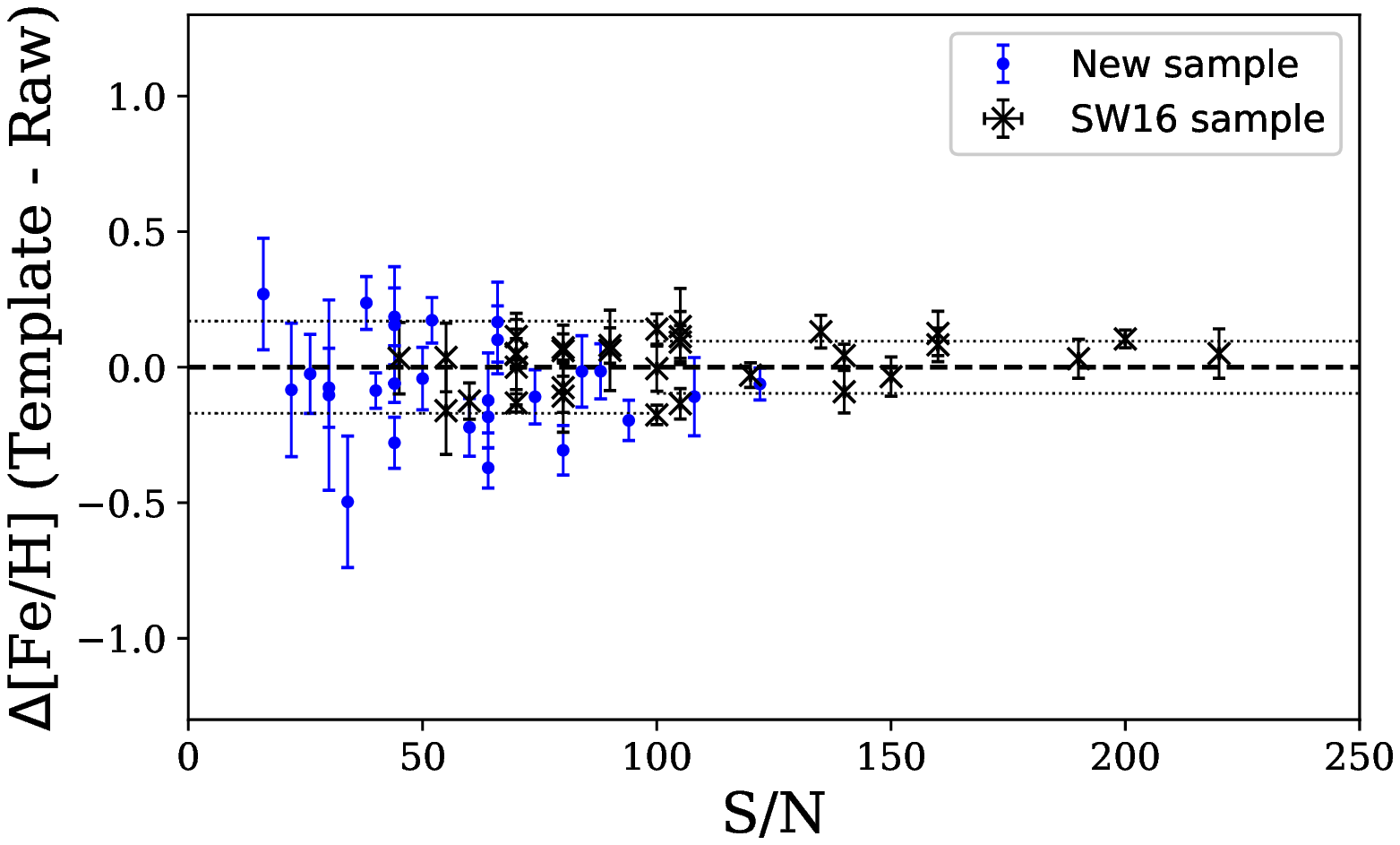}\label{fig:Raw_vs_Tempa}}
\subfigure[Pseduo EWs]{\includegraphics[scale=0.53,trim=0in 0in 0.45in 0.45in,clip]{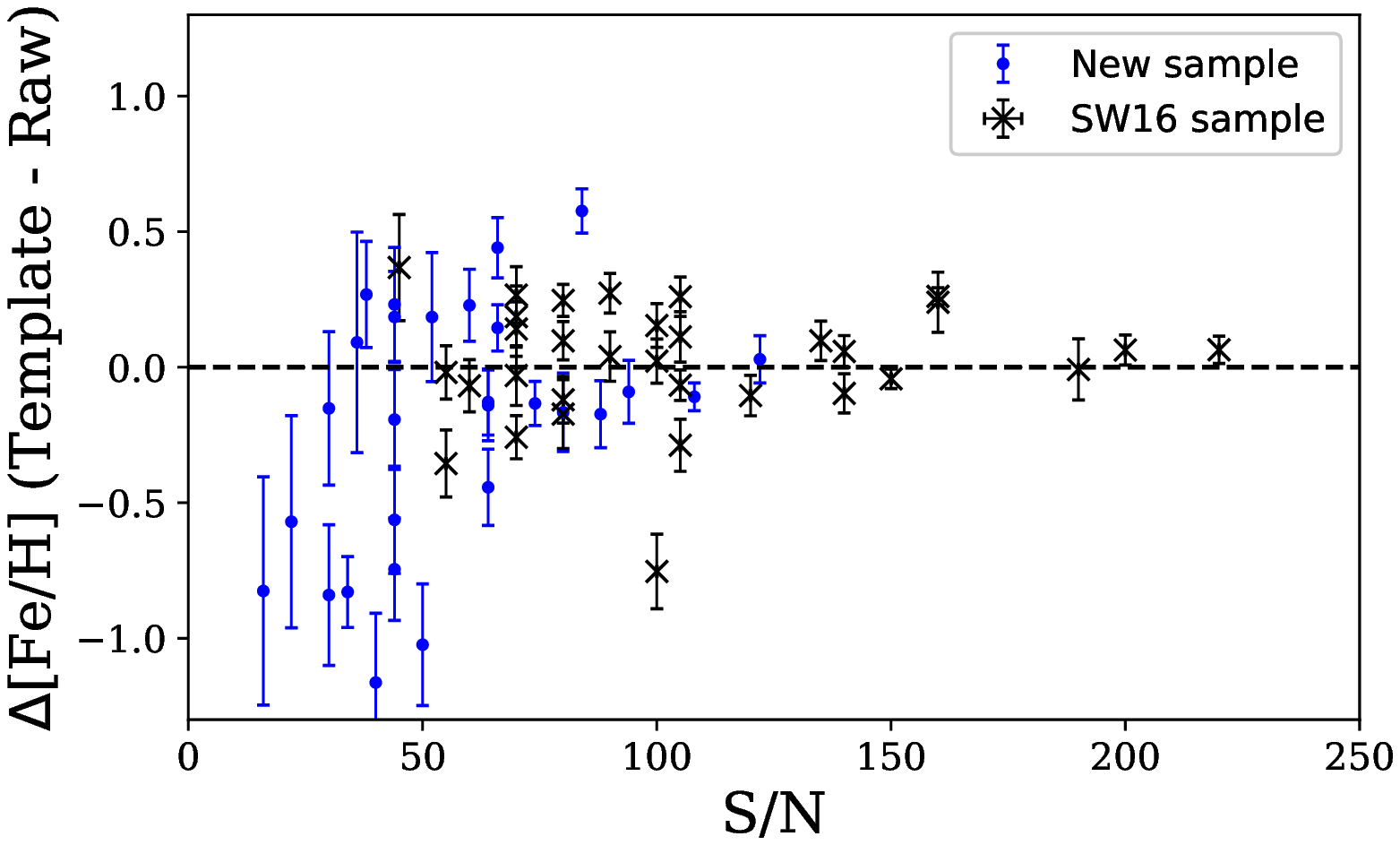}\label{fig:Raw_vs_Tempb}}
\caption{Offsets between [Fe/H] determinations on template-fitted vs. raw spectra, for Voigt profile fits (left) and pseudo EWs (right), as a function of the S/N of the spectrum (per resolution element).  The black crosses are the calibration clusters from \citet{SakWall2016} while the blue circles are the new targets from this paper.  The dashed line shows equal agreement.  The dotted lines in the left panel show the standard deviation of the offsets for low S/N (with S/N$<100$) and high S/N ($\ge100$) GCs.\label{fig:Raw_vs_Temp}}
\end{center}
\end{figure*}

This section assesses whether the template-fitting process accurately matches the true spectra of the GCs.  As mentioned previously, the raw spectra often have severe sky line contamination, particularly for the first and third CaT lines.  For that reason, only the second CaT line is used for this comparison.  (Section \ref{subsec:Voigt_vs_Int} will also demonstrate that using only the second CaT line leads to a stronger relationship with [Fe/H].)  Figure \ref{fig:Raw_vs_Temp} shows comparisons between the [Fe/H] ratios derived from the template-fitted vs. raw spectra, as a function of the S/N ratio of the spectrum.  This figure demonstrates that the templates generally do an excellent job of matching the second CaT line, although the disagreement increases with lower S/N.  Visual inspection of the Voigt profile fits to the low S/N raw spectra show that this disagreement is generally driven by noise or sky features that alter the line profile, such that the Voigt fits do not accurately represent the line in the raw spectra.  These effects fluctuate from spectrum to spectrum, such that the Voigt fits to the raw and template-fitted spectra agree on average (Figure \ref{fig:Raw_vs_Tempa}).  The pseudo EWs of the raw spectra, on the other hand, do not match well at low S/N.  Again, visual inspection suggests that this is a result of residual sky lines in the bandpass definitions (see Section \ref{subsec:Integrations}).  In very low S/N spectra, the residual sky lines subtract flux from the line strength, leading to lower metallicities.  Ultimately, it appears that the template-fitted spectra do an excellent job representing the raw spectra, and can better capture the strength of the second CaT line at low S/N.

However, the templates likely will not perfectly match the true spectra, particularly for spectra with low S/N, and there will be some uncertainty in [Fe/H] associated with the template fit.  This uncertainty can be estimated as a function of S/N, using the dispersion in Figure \ref{fig:Raw_vs_Tempa}.  The dotted lines show the standard deviations of the [Fe/H] differences, with the GCs split into two samples: low S/N GCs, with S/N$<100$, and high S/N GCs, with S/N$\ge 100$.  The resulting standard deviations are $0.2$~dex for the low S/N GCs, and $0.1$~dex for the high S/N GCs.\footnote{Note that the offsets are nearly identical if three groups are used instead of two.} These standard deviations are used in Section \ref{subsec:FeHresults} to quantify the uncertainty in [Fe/H] that arises from the template-fitting process.

\subsection{Voigt Profile Fits vs. Pseudo EWs}\label{subsec:Voigt_vs_Int}
Now that the template-fitted spectra have been selected, the two measurement techniques (pseudo EW integrations and Voigt profile fits) are compared.  The adopted [Fe/H] vs. line strength calibrations with the template-fitted spectra from \citet{SakWall2016} are presented in Table \ref{table:Calibrations}.  Here the relationships with the strengths of the second CaT line and with the sum of the second and third CaT lines are both given, along with the resulting $\chi^2$ values for the fits.  These exact $\chi^2$ values are particularly sensitive to the adopted uncertainties in [Fe/H]---this will be discussed in more detail in Section \ref{subsec:CaH}.  This section is more concerned with the differences between the various ways of quantifying the CaT strength.

\begin{figure}
\begin{center}
\centering
\hspace*{-0.15in}
\includegraphics[scale=0.55,trim=0.15in 0.0in 0.5in 0.3in,clip]{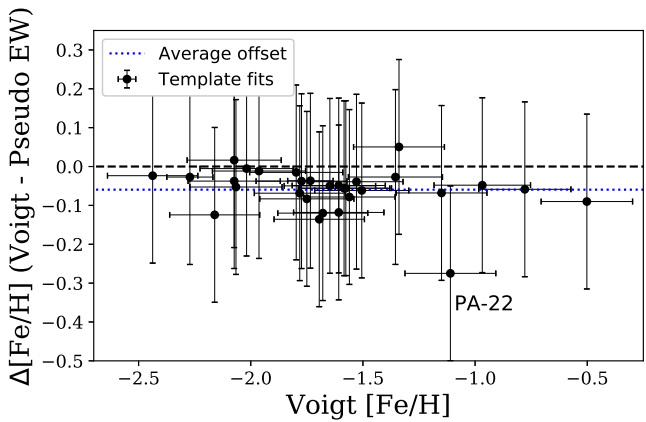}
\caption{Differences in the resulting [Fe/H] between Voigt profile fits and pseudo EWs in template-fitted spectra.  Each point is a different GC; the error bars are the total errors (see Section \ref{subsec:Raw_vs_Temp}).  The dotted blue line shows the average offset of -0.06.}\label{fig:Voigt_vs_PEW}
\end{center}
\end{figure}

There are a couple of key findings in Table \ref{table:Calibrations}. First, Table \ref{table:Calibrations} shows that the $\chi^2$ values are lower when only the second CaT line is used. Second, there are minimal differences between the relationships for the pseudo-EWs and the Voigt profile fits, although the $\chi^2$ values are slightly smaller for the Voigt profile fits on template-fitted spectra.  Finally, the relationships in Table \ref{table:Calibrations} are also very similar to those in the original \citet{SakWall2016} analysis, suggesting that the new measurement techniques adopted here have not significantly altered the CaT line strengths.  The slight differences in the calibrations lead to very small offsets in the derived [Fe/H] ratios.  Figure \ref{fig:Voigt_vs_PEW} shows the differences in [Fe/H] for Voigt profile fits vs. pseudo-EWs in template-fitted spectra; on average, the differences are -0.06.  There is some hint of a trend with [Fe/H], where metal-rich GCs are found to be more metal-rich with pseudo EW integrations; this may reflect that the pseudo EW measurements are sensitive to the presence of nearby spectral lines, which are masked out in the Voigt profile fits---this does seem to be the case for PA~22, the most notable outlier in Figure \ref{fig:Voigt_vs_PEW}.

\begin{figure*}
\begin{center}
\centering
\subfigure[{[Fe/H]} calibration]{\includegraphics[scale=0.53,trim=0in -0.1in 0.45in 0.5in,clip]{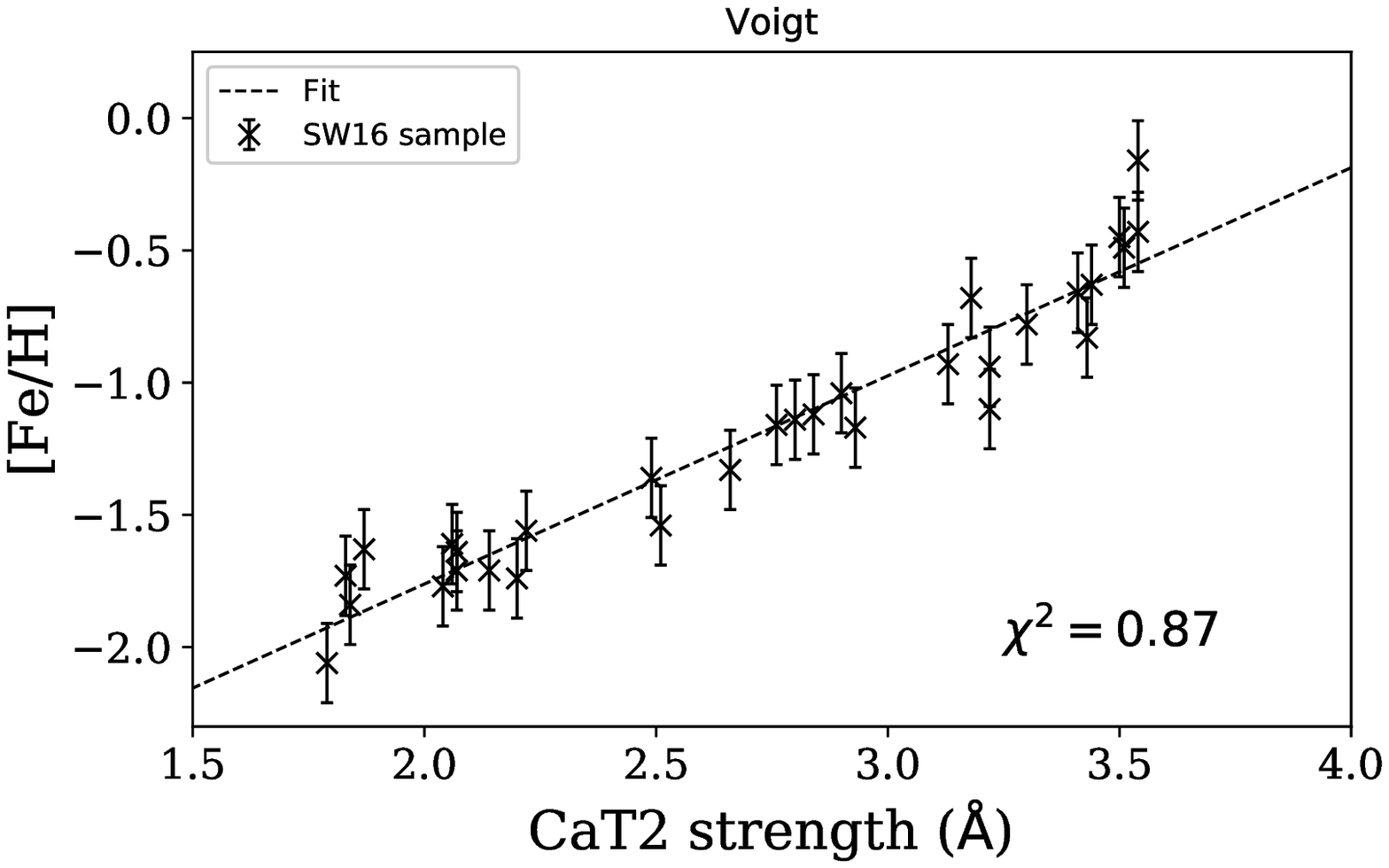}\label{fig:FeHcalibration}}
\subfigure[{[Ca/H]} calibration]{\includegraphics[scale=0.53,trim=0in -0.1in 0.45in 0.5in,clip]{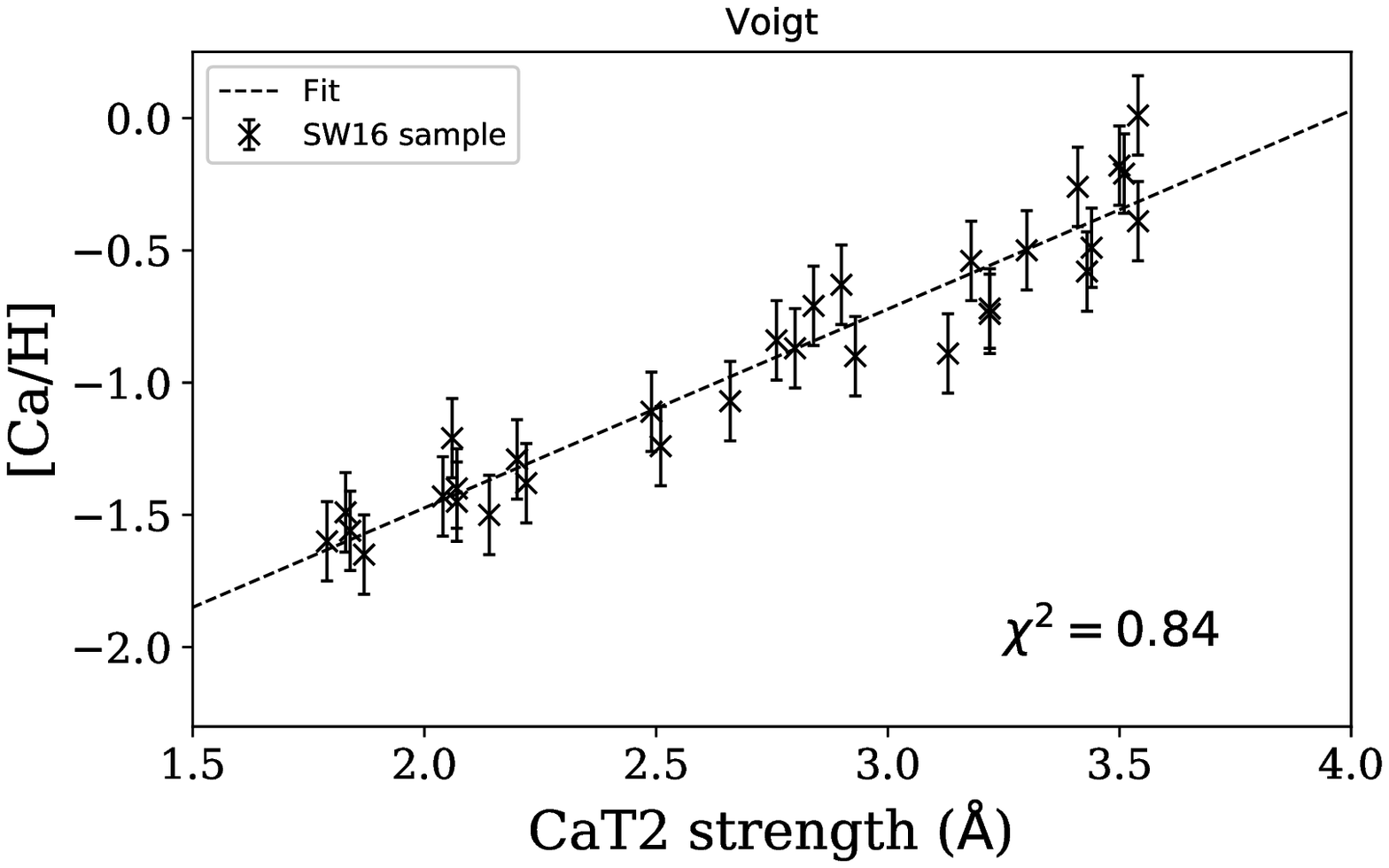}\label{fig:CaHcalibration}}
\caption{Relationships between GC [Fe/H] (left) and [Ca/H] (right) versus the strength of the second IL CaT line, as determined by Voigt fits to template-fitted spectra.  The crosses show the M31 GCs from \citet{SakWall2016}, whose [Fe/H] ratios were determined at high spectral resolution \citep{Colucci2014,Sakari2015,Sakari2016}.  The dashed lines show the linear fits to the points; the resulting $\chi^2$ value is given in the lower right corner.\label{fig:Calibrations}}
\end{center}
\end{figure*}

Ultimately, the best relationship is obtained with Voigt profile fits to the second CaT line.  Although the $\chi^2$ fits are only marginally lower with Voigt profile fits, compared to the pseudo EWs, Figure \ref{fig:Raw_vs_Temp} indicates that the Voigt profile fits may be reliable for lower S/N spectra.  Since the new targets in this paper extend to lower S/N ratios than the original calibration sample from \citet{SakWall2016}, the Voigt profile fits are ultimately adopted.  The final relationship between the strength of the second CaT line and the GC [Fe/H] is shown in Figure \ref{fig:FeHcalibration}.

\begin{table}
\centering
\begin{center}
\caption{CaT strength vs. metallicity relationships, for different
  indicators of the CaT strength, as derived from template-fitted spectra.\label{table:Calibrations}}
  \begin{tabular}{@{}lllc@{}}
  \hline
& \multicolumn{1}{l}{Measurement} & & \\
Line(s) & \multicolumn{1}{l}{Technique}  & Relation &  $\chi^{2a}$   \\
\hline
[Fe/H] fits  & & \\
CaT2 & Pseudo EWs & $[\rm{Fe/H}] = 0.77\times \rm{CaT} - 3.33$ & 0.88\\
     & Voigt Fits & $[\rm{Fe/H}] = 0.79\times \rm{CaT} - 3.33$ & 0.87 \\
 & & & \\
CaT2+3 & Pseudo EWs & $[\rm{Fe/H}] = 0.46\times \rm{CaT} -3.49$ & 0.98\\
     & Voigt Fits & $[\rm{Fe/H}] = 0.47\times \rm{CaT} - 3.49$ & 0.99 \\
 & & & \\
 & & & \\
$[$Ca/H$]$ fits & & \\
CaT2 & Pseudo EWs & $[\rm{Ca/H}] = 0.73\times \rm{CaT} - 2.96$ & 0.91 \\
     & Voigt Fits & $[\rm{Ca/H}] = 0.75\times \rm{CaT} - 2.98$ & 0.84 \\
\hline
\end{tabular}
\end{center}
\raggedright $^{a}$ Note that the $\chi^2$ values are very sensitive to the adopted uncertainties; see Section \ref{subsec:CaH}.
\end{table}

\subsection{[Ca/H] vs. [Fe/H]}\label{subsec:CaH}
Since the original CaT calibration paper by \citealt{AZ88}, the CaT lines have been used to determine a GC's overall metallicity, which is often quantified as [Fe/H].  However, recent papers (e.g., \new{\citealt{DaCosta2016}},\citealt{Usher2019}) have suggested that the strengths of the CaT lines are more sensitive to a GC's [Ca/H] than to [Fe/H].  Although the CaT lines are caused by \ion{Ca}{2} atoms, the strengths of the strong CaT lines are primarily driven by electron pressure broadening, which is sensitive to the overall number of atoms in the stellar atmospheres (see, e.g., \citealt{Battaglia2008}).  Iron is generally more abundant than Ca, meaning that the Fe abundance likely has a greater influence on the CaT lines.  However, there are other elements that do not always track Fe and that may be more abundant, including Mg.  Mg, one of the $\alpha$-element, does not also track iron because the two elements have different nucleosynthetic sites (e.g., \citealt{WoosleyWeaver1995}).  Ca is also often considered to be an $\alpha$-element, so even though it is less abundant than Fe the [Ca/H] ratio may better trace a GC's overall metallicity.\footnote{However, note that Mg also has a different nucleosynthetic site than Ca, meaning that the two do not always track each other (e.g., \citealt{McW2013}).}  For these reasons, \citet{Usher2019} investigated trends in IL CaT strength with [Fe/H] and [Ca/H] for GCs in the Milky Way and its neighboring satellites.  They found that the IL CaT line strength had a more significant correlation with [Ca/H] than [Fe/H].  
These findings were also supported by the previous work by \citet{SakWall2016}, who found that a GC's [Ca/Fe] ratio did seem to affect its CaT strength.

To test whether the CaT line strengths of the M31 GCs from \citet{SakWall2016} are more correlated with [Ca/H] or [Fe/H], the calibrations from Section \ref{subsec:Voigt_vs_Int} were redone with [Ca/H] instead of [Fe/H].  The [Ca/H] ratios were calculated from the [Fe/H] and [Ca/Fe] ratios quoted in \citet{Colucci2009,Colucci2014} and \citet{Sakari2015,Sakari2016}.  The resulting relationships are listed in Table \ref{table:Calibrations} and the relationship with the Voigt profiles is shown in Figure \ref{fig:CaHcalibration}.

As mentioned in Section \ref{subsec:Voigt_vs_Int}, the goodness of the fit (as quantified by $\chi^2$) is very sensitive to the adopted uncertainties in [Fe/H] and [Ca/H].  The values for each cluster that are quoted in \citet{Colucci2009,Colucci2014} and \citet{Sakari2015,Sakari2016} are typically {\it random} errors; however, a comparison like this requires taking the systematic errors into account.  \citet{Sakari2014} present a detailed discussion of the potential systematic errors that can occur in high-resolution IL analyses.  To summarize, they generally find that systematic uncertainties in [Ca/Fe] are $<$0.05 dex, while the systematic uncertainties in [Fe/H] can be larger, at 0.1-0.2 dex.\footnote{Note that the systematic uncertainties can be larger in extreme cases; see \citet{Sakari2014}.}  The lower offsets in [Ca/Fe] indicate that [Fe/H] and [Ca/H] change together.  Without going through a detailed systematic analysis, here a constant systematic offset of 0.15 dex is adopted for both [Fe/H] and [Ca/H] for all GCs.  Table \ref{table:Calibrations} and Figure \ref{fig:Calibrations} show that the [Ca/H] relationship has a slightly lower $\chi^2$ for the Voigt fits (though not the pseudo EWs), with a similar relationship with CaT strength as [Fe/H].  Although the [Ca/H] relationships yield slightly lower $\chi^2$ values, the differences are negligible.  For simplicity and easier comparisons with the literature, [Fe/H] is adopted in this paper.  However, it is worth noting that a GC's $\alpha$-abundances {\it can} affect the CaT-based [Fe/H]; this will be discussed further below.

\subsection{Results from Template-Fitted Spectra and Comparisons with the Literature}\label{subsec:FeHresults}
The derived [Fe/H] ratios for the new M31 GCs are shown in Table \ref{table:FeH}.  The uncertainties from the Voigt profile fits, including Monte Carlo sampling, and the assumed uncertainty from template fitting (Section \ref{subsec:Raw_vs_Temp}) are listed separately; the final adopted uncertainty in [Fe/H] is the two values added in quadrature.

\begin{table}
\centering
\begin{center}
\caption{CaT-based [Fe/H] ratios.\label{table:FeH}}
\newcolumntype{d}[1]{D{,}{\;\pm\;}{#1}}
\begin{tabular}{@{}lcccc@{}}
  \hline
        &        & $\sigma$[Fe/H] & $\sigma$[Fe/H] & $\sigma$[Fe/H] \\
Cluster & [Fe/H] & Random$^{a}$ & Template$^{b}$ & Total$^{c}$\\
  \hline
PA-01   & -2.07 & 0.05 & 0.20 & 0.21 \\
PA-04   & -2.07 & 0.06 & 0.20 & 0.21 \\
PA-09   & -1.56 & 0.04 & 0.20 & 0.20 \\
PA-11   & -2.16 & 0.02 & 0.20 & 0.20 \\
PA-14   & -1.61 & 0.02 & 0.20 & 0.20 \\
PA-16   & -2.44 & 0.02 & 0.20 & 0.20 \\
H4      & -1.77 & 0.03 & 0.20 & 0.20 \\
H5      & -1.65 & 0.04 & 0.20 & 0.20 \\
H7      & -1.34 & 0.03 & 0.20 & 0.20 \\
PA-22   & -1.11 & 0.03 & 0.20 & 0.20 \\
PA-27   & -1.51 & 0.06 & 0.20 & 0.21 \\
dTZZ-05 & -0.78 & 0.05 & 0.20 & 0.21 \\
H11     & -1.53 & 0.06 & 0.20 & 0.21 \\
H12     & -1.73 & 0.02 & 0.10 & 0.10 \\
H18     & -1.80 & 0.06 & 0.20 & 0.21 \\
H19     & -1.78 & 0.03 & 0.20 & 0.20 \\
PA-36   & -0.97 & 0.08 & 0.20 & 0.22 \\
G339    & -1.75 & 0.06 & 0.20 & 0.21 \\
PA-37   & -0.50 & 0.04 & 0.20 & 0.20 \\
H22     & -1.68 & 0.02 & 0.20 & 0.20 \\
EXT8    & -2.27 & 0.02 & 0.10 & 0.10 \\
PA-41   & -1.96 & 0.06 & 0.20 & 0.21 \\
H24     & -1.58 & 0.04 & 0.20 & 0.20 \\
PA-44   & -2.07 & 0.05 & 0.20 & 0.21 \\
PA-46   & -2.02 & 0.06 & 0.20 & 0.21 \\
H25     & -1.58 & 0.04 & 0.20 & 0.20 \\
B517    & -1.36 & 0.06 & 0.20 & 0.21 \\
H27     & -1.61 & 0.06 & 0.20 & 0.21 \\
PA-52   & -1.15 & 0.03 & 0.20 & 0.20 \\
dTZZ-21 & -1.69 & 0.02 & 0.20 & 0.20 \\
\hline
\end{tabular}
\end{center}
\medskip
\raggedright
$^{a}$ The uncertainty in [Fe/H] due to the Voigt profile fit.\\
$^{b}$ The adopted uncertainty in [Fe/H] that results from the template fitting, as determined by the S/N ratio of the CaT spectrum (see Section \ref{subsec:Raw_vs_Temp}).\\
$^{c}$ The total error, which is $\sigma$[Fe/H]$_{\rm{rand}}$ and $\sigma$[Fe/H]$_{\rm{temp}}$ added in quadrature.\\
\end{table}

\begin{figure*}
\begin{center}
\centering
\hspace*{-0.15in}
\includegraphics[scale=0.65,trim=0in 0.0in 0in 0.in,clip]{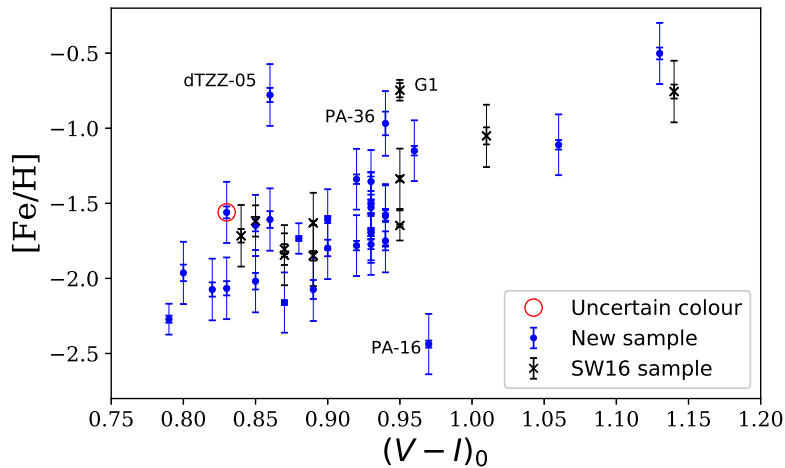}
\caption{A comparison of the [Fe/H] ratio from this work with the integrated $(V-I)_0$ colours quoted in \citet{Mackey2019}.  The red circled point is PA-09, which has an uncertain colour.  In general there is a trend between increasing [Fe/H] and redder colours; the interesting outliers are labeled, and are discussed in the text.}\label{fig:ColorTest}
\end{center}
\end{figure*}

All of these GCs have de-reddened IL photometry from \citet{Huxor2014}, as quoted in \citet{Mackey2019}.  A GC's IL $(V-I)_0$ colour is known to correlate with the metallicity, albeit in a complicated way (e.g., \new{\citealt{Yoon2006}}).  Figure \ref{fig:ColorTest} shows the CaT-based [Fe/H] compared with the IL $(V-I)_0$ colour.  Note that one cluster, PA-09, has an uncertain colour due to the presence of a nearby bright star; this point is circled in Figure \ref{fig:ColorTest}.  This figure demonstrates that metal-rich GCs are generally redder, while metal-poor GCs are generally bluer.  However, several intriguing targets stand out in this plot.  They are labelled and will be discussed further below.

\begin{figure*}
\begin{center}
\centering
\hspace*{-0.15in}
\subfigure[]{\includegraphics[scale=0.53,trim=0in -0.1in 0.45in 0.2in,clip]{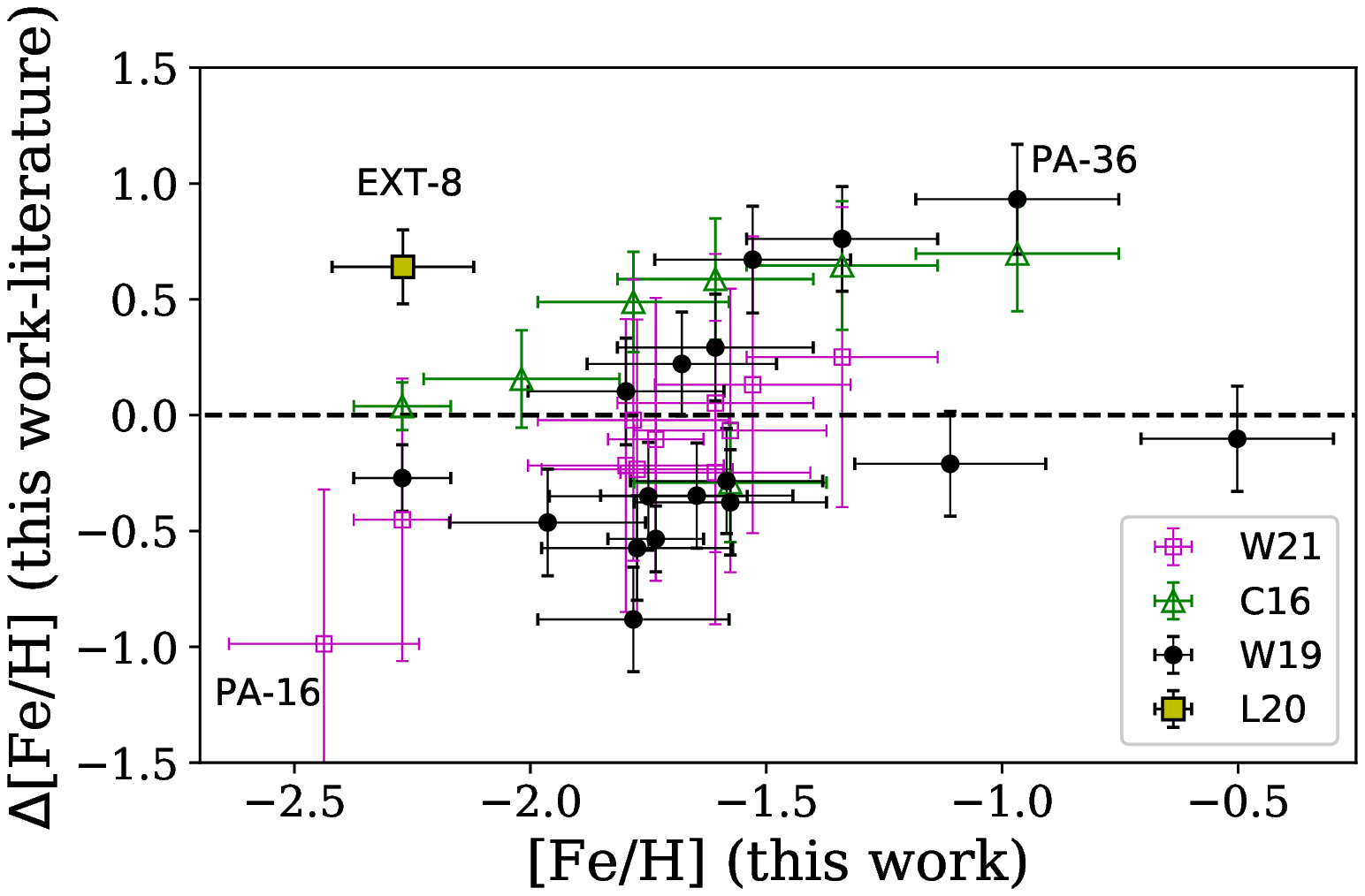}\label{fig:CompFeH}}
\subfigure[]{\includegraphics[scale=0.53,trim=0in -0.1in 0.45in 0.2in,clip]{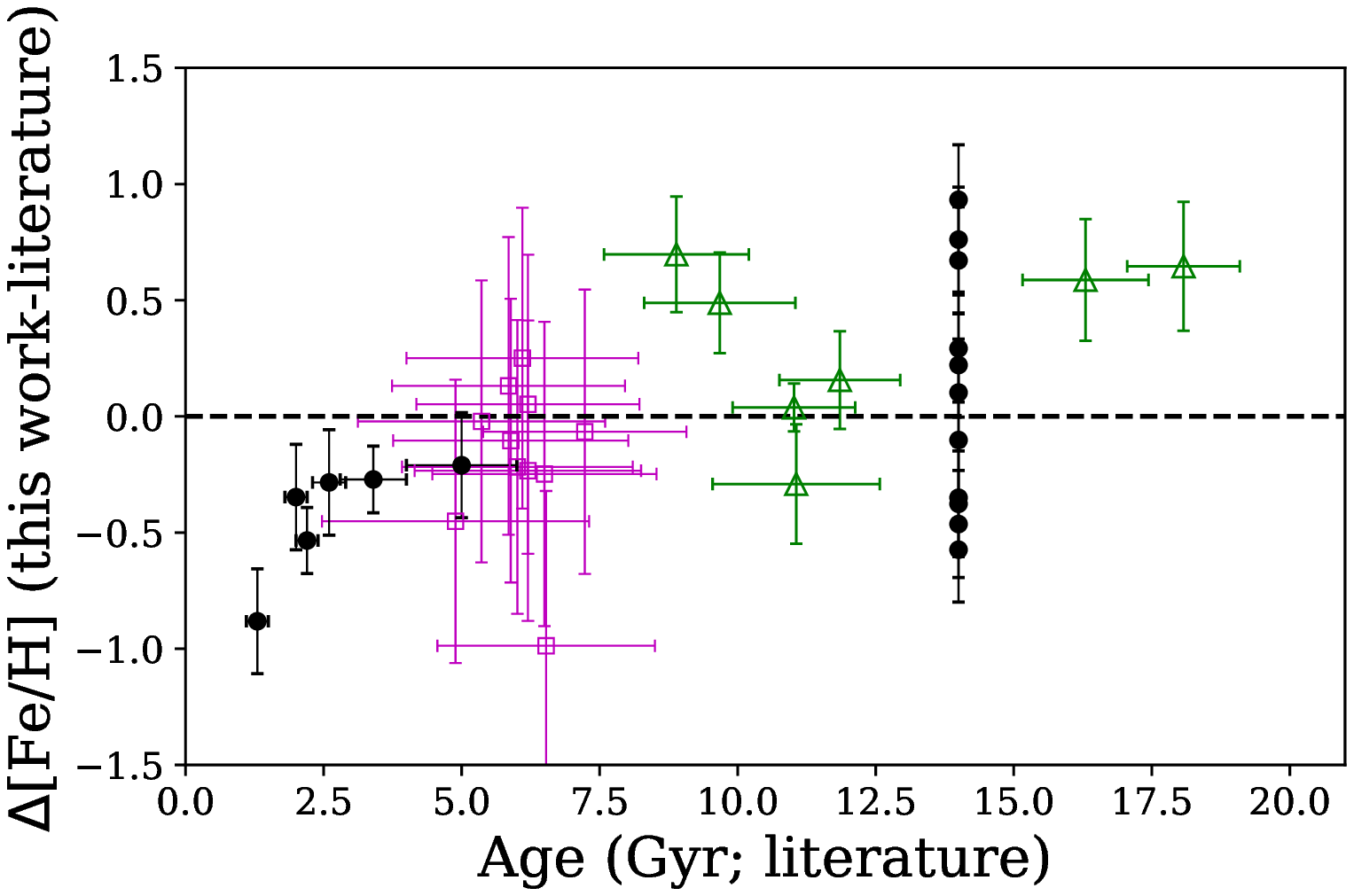}\label{fig:CompFeH_age}}
\caption{Differences in [Fe/H] (this work $-$ literature values) between the CaT values from this work and previous results from \citet[C16; green crosses]{Chen2016}, \citet[W19; open magenta squares]{Wang2019}, \citet[W21; black circles]{Wang2021}, and \citet[L20; yellow square]{Larsen2020}.  {\it Left:} $\Delta$[Fe/H] as a function of the CaT-based [Fe/H]. {\it Right: } $\Delta$[Fe/H] as a function of the age from the literature.  Note that some clusters are included in multiple studies.  The \citet{Chen2016} results are averages of their results with two different models.}\label{fig:CompLit}
\end{center}
\end{figure*}

Several of these clusters have been previously observed and analyzed.  Figure \ref{fig:CompLit} compares the [Fe/H] ratios from this work to those from the following papers.
\begin{itemize}
    \item \citet{Chen2016} analyzed optical IL spectra from the Large Sky Area Multi-Object Fiber Spectroscopic Telescope (LAMOST) survey for 7 of the GCs in this paper, using simple stellar population fits to derive ages and metallicities.  Note that \citet{Chen2016} present multiple values with different models.  In Figure \ref{fig:Comp} the values from two models were averaged, since the values generally are not significantly different.
    \item \citet{Wang2019} used IL photometry and stellar population synthesis to determine ages and metallicities for 17 of the GCs in this study.
    \item \citet{Wang2021} analyzed additional LAMOST IL spectra and photometry, deriving ages and metallicities for 11 of the GCs in this paper.
    \item \citet{Larsen2020} analyzed a high-resolution IL spectrum of EXT~8, while \citet{Larsen2021} obtained resolved {\it Hubble Space Telescope} photometry of EXT~8.
\end{itemize}
Note that some clusters are analyzed in multiple papers.  EXT~8, for example, was analyzed in all four samples.  Figure \ref{fig:CompFeH} compares the $\Delta$[Fe/H] ratios as a function of the [Fe/H] ratio from this paper, while Figure \ref{fig:CompFeH_age} shows $\Delta$[Fe/H] as a function of the ages from C16, W19, and W21.  Interesting clusters are labeled in Figure \ref{fig:CompFeH}.  This comparison shows considerable variations between the CaT-based [Fe/H] ratios from this paper and those in the literature (as well as large variations amongst the previous papers).

Comparisons between the results in this paper and those in the literature for specific GCs are discussed below.

\subsubsection{EXT~8}\label{subsubsec:EXT8}
All four groups analyzed EXT~8, with large variations in its [Fe/H], as seen in Figure \ref{fig:CompFeH}.  \citet{Larsen2020} quote $[\rm{Fe/H}]~=~-2.91\pm0.04$, which would make EXT~8 the lowest metallicity GC ever found.  From resolved {\it HST} photometry, \citet{Larsen2021} find that EXT~8 is consistent with being as or more metal-poor than the Milky Way GC M15 (which \citealt{Sakari2013} find to have $[\rm{Fe/H}]~=~-2.3$, from high-resolution IL spectroscopy).  The metallicities for EXT~8 from the other papers range from $[\rm{Fe/H}]~=~-1.8$ \citep{Wang2021} to $[\rm{Fe/H}]~=~-2.3$ \citep{Chen2016}.  The value from this paper, $[\rm{Fe/H}]~=~-2.27$, is higher than the high-resolution value from \citet{Larsen2020}, but is generally consistent with the other IL studies. 

This discrepancy with the high-resolution study may be a result of the lack of metal-poor stars in the template set.  Indeed, the strength of the second CaT line is slightly weaker in the raw spectrum than in the template-fitted spectrum.  However, it is worth noting that in order to obtain the \citet{Larsen2020} [Fe/H] ratio from the second CaT, according to the relationship in Table \ref{table:Calibrations}, the CaT line would have to be 0.53 \AA, over 0.5 \myAA weaker than its template-fitted value.  Instead, the linear relationship between CaT strength and IL [Fe/H] may break down at the low-metallicity end, similar to individual stars \citep{Starkenburg2010}.  There may be a more complicated relationship between CaT strength and metallicity; however, there are not currently enough very metal-poor GCs to perform this calibration.

\subsubsection{Young GCs?}\label{subsubsec:Young}
Several GCs have been previously found to be intermediate-age or young, with ages $<5$ Gyr, by \citet{Wang2019} and \citet{Wang2021}.  \new{The simulations from \citet{Hughes2019} suggest that younger outer halo GCs trace accretion from more massive satellites.}  However, Figure \ref{fig:CompFeH_age} demonstrates that, although there is large dispersion for older GCs, the GCs that were previously found to be young all have lower CaT-based [Fe/H] ratios, compared to those derived in the literature.  This difference may be a result of the age-metallicity degeneracy that complicates optical analyses.  As a result of this degeneracy, optical IL spectra of metal-poor GCs may be well fit by IL spectra of metal-rich, young GCs.  The CaT lines are less sensitive to cluster age, at least for clusters older than $\sim1$ Gyr \citep{Usher2019}, meaning that the CaT lines may help break the age-metallicity degeneracy \new{(e.g., \citealt{Usher2019b})}.

\subsubsection{PA-16}\label{subsubsec:PA16}
\begin{figure*}
\begin{center}
\centering
\hspace*{-0.15in}
\subfigure[]{\includegraphics[scale=0.53,trim=0in -0.1in 0.45in 0.2in,clip]{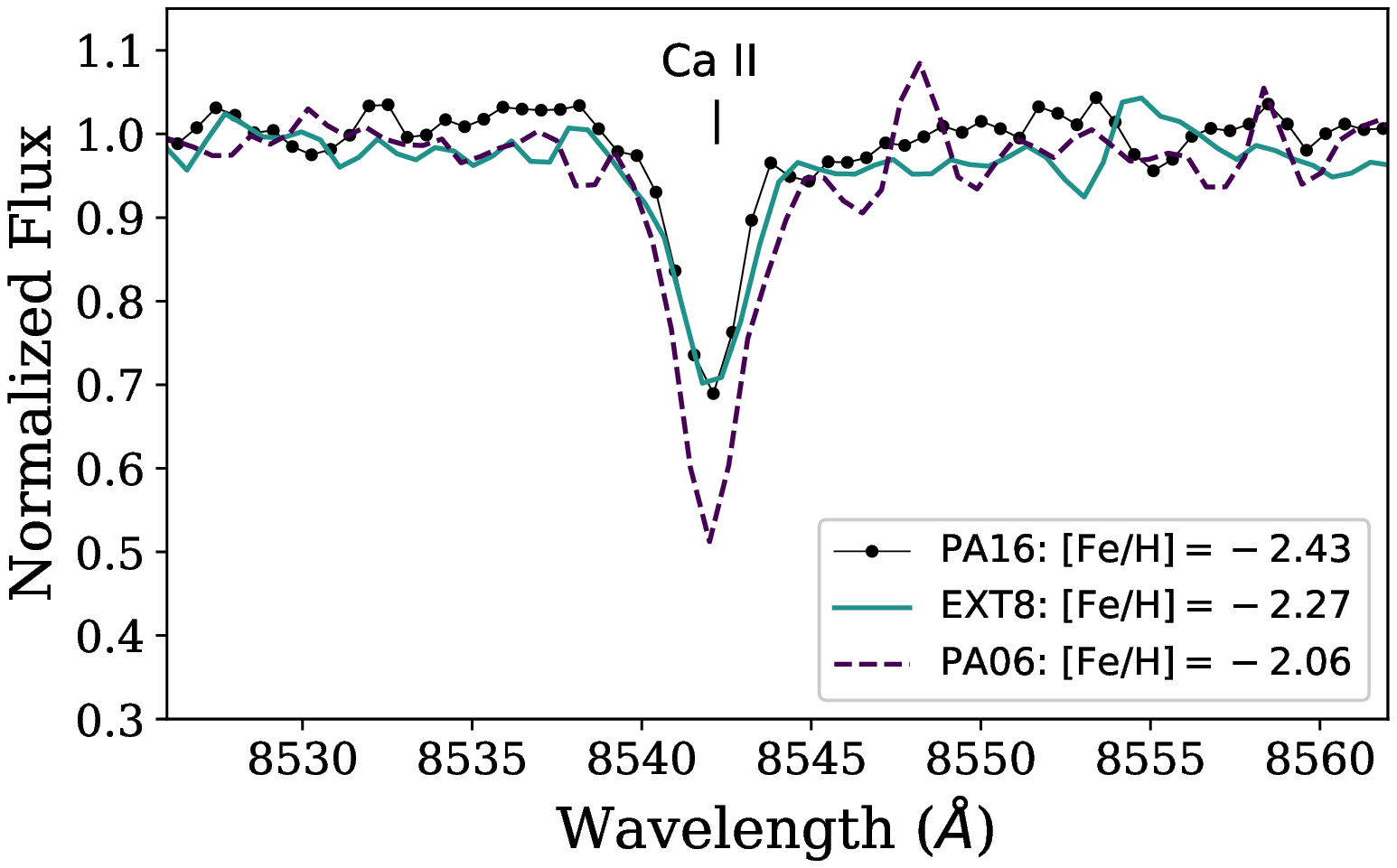}\label{fig:PA16}}
\subfigure[]{\includegraphics[scale=0.53,trim=0in -0.1in 0.45in 0.2in,clip]{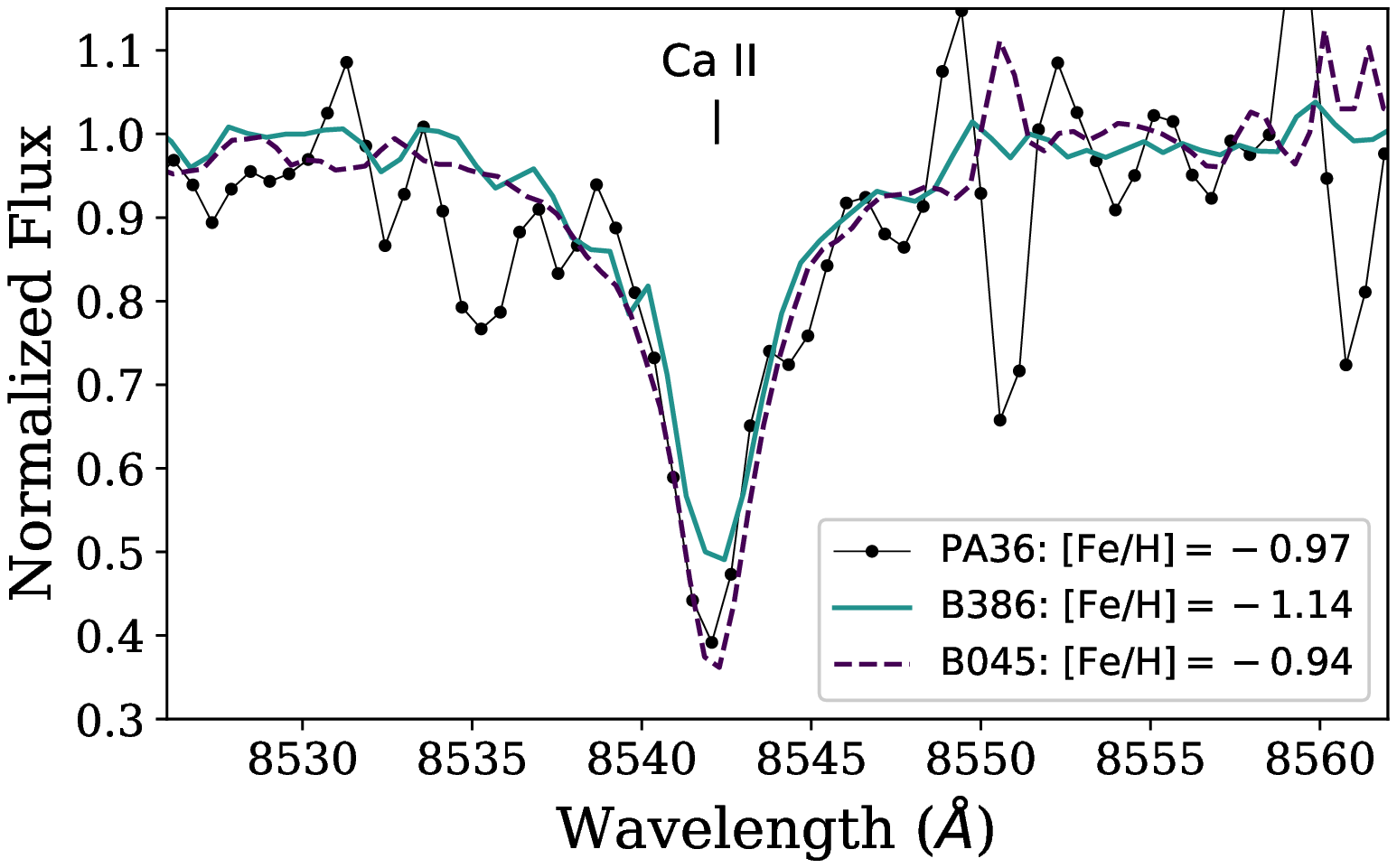}\label{fig:PA36}}
\caption{Comparisons of raw IL spectra of the second CaT line.  {\it Left: } The GC PA-16, which is found to be very metal-poor in this work, is compared with two other metal-poor GCs, EXT~8 and PA-06.  {\it Right: } PA-36, which is found to be fairly metal-rich, is compared with B386 and B405.}\label{fig:CompPlotCaT}
\end{center}
\end{figure*}

PA-16 is a clear outlier in both Figure \ref{fig:ColorTest} and Figure \ref{fig:CompFeH}.  The CaT-based metallicity, $[\rm{Fe/H}] = -2.43$, is 1 dex more metal-poor than the value from \citet{Wang2021}.  Similarly, the CaT-based metallicity is lower than expected from its $(V-I)_0$ color.  Figure \ref{fig:PA16} shows the second IL CaT line, along with two other metal-poor GCs, EXT-8 and PA-06 (from \citealt{SakWall2016}).  The strength of PA-16's second CaT line is similar to that of EXT-8, hinting that PA-16 is indeed a metal-poor cluster.

There are several options to explain why PA-16's metallicity is so much lower than expected.  First, one option is that another object was observed instead of PA-16.  However, PA-16's radial velocity, $v_{\rm{helio}} = -469.5\pm20$~km/s, agrees reasonably well with the value from \citet{Veljanoski2014}, $-490\pm15$~km/s; at PA-16's location, it is unlikely that a Milky Way foreground star would have such a large negative velocity.  Another possibility is that something could have gone wrong with the data reduction, though the other GC observed that night, PA-04, does not seem to have a similar issue.  Higher values of reddening could somewhat explain the discrepancy in Figure \ref{fig:ColorTest}.  

\begin{figure*}
\begin{center}
\centering
\includegraphics[scale=0.58,trim=0.8in 0.in 1.0in 0.3in,clip]{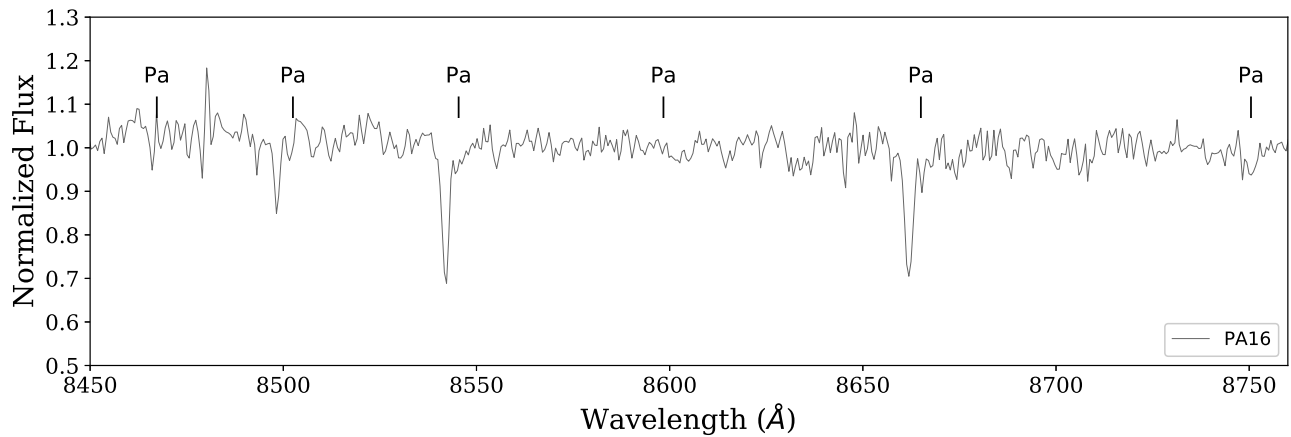}
\caption{Locations of potential hydrogen Paschen lines in the GC PA-16.}\label{fig:CompPlotPaT}
\end{center}
\end{figure*}

Finally, it is also possible that this discrepancy could reflect something physical about the cluster, e.g., its age.  If PA~16 is very young, in principle its weaker IL CaT lines could be caused by contributions from hotter main sequence stars.  Young clusters are also expected to show hydrogen Paschen lines; \citet{Usher2019} show in their sample of IL spectra of nearby GCs that Paschen lines become detectable in IL spectra for GCs younger than $\sim200$ Myr, although they are relatively weak; \new{they further find that the Paschen lines may affect the CaT in GCs as old as $\sim 2$ Gyr.\footnote{\new{\citet{Usher2019} also find that the CaT lines may not reliably track metallicity in GCs with ages around 2 Gyr, due to contributions from asymptotic giant branch stars.}}}  Figure \ref{fig:CompPlotPaT} highlights the expected locations of Paschen lines.  There are {\it possible} hints of Paschen lines in the PA-16 spectrum, particularly in the wings of the second CaT line.  \new{However, it is difficult to explain how a very young GC could end up in the outer halo.  It is also worth noting that \citet{Wang2021} identified PA-16 as being intermediate-aged.}

Another possibility is that PA-16 has a significant population of extremely blue HB stars, which could produce Paschen lines and weaken the IL CaT lines.  Both young and HB stars should have less of an effect on the CaT lines than on spectra further in the blue.  Finally, another option is that PA-16 is a GC with very low [$\alpha$/Fe].  Since the strength of the CaT lines correlates with overall metallicity, a very low [$\alpha$/Fe] would mean that its [Fe/H] would be higher than predicted from its CaT strength. Further observations of PA-16 are needed to determine its age and chemical composition.

\subsubsection{PA-36}\label{subsubsec:PA16}
This analysis finds PA-36 to be more metal-rich than found by \citet{Chen2016} and \citet{Wang2019}, as shown in Figure \ref{fig:CompLit}.  PA-36's CaT-based metallicity is also slightly higher than predicted from its $(V-I)_0$ color (Figure \ref{fig:ColorTest}).  Figure \ref{fig:PA36} shows the second CaT line in the PA-36 spectrum, along with two other M31 GCs, B045 and B386, from \citet{SakWall2016}.  The strength and shape of PA-36's second CaT line is very similar to B045's, indicating that they have similar metallicities.  

However, it is worth highlighting another GC in Figure \ref{fig:ColorTest}, which also has a higher CaT-based [Fe/H] than predicted from its IL $(V-I)_0$ colour: the massive GC, G1.  In G1's case, its bluer colours may be a result of a large helium enhancement and the presence of extreme blue horizontal branch stars 
\citep{Nardiello2019}.  G1 also has a suspected Fe spread \citep{Meylan2001,Nardiello2019}, which influences its single IL [Fe/H] (Sakari et al., {\it in prep.}).  Indeed, \citet{Sakari2021} found a high-resolution, optical [Fe/H] ratio that was lower than the CaT-based [Fe/H].  The presence of abundance spreads, whether from He or Fe, could potentially explain the offsets between PA-36 and the literature.  \new{Alternatively, PA-36's bluer $(V-I)_0$ colours could indicate that it is younger than the other GCs.}

\subsubsection{dTZZ-05}\label{subsubsec:dTZZ-05}
\begin{figure}
\begin{center}
\centering
\hspace*{-0.15in}
\includegraphics[scale=0.53,trim=0in 0.0in 0in 0.in,clip]{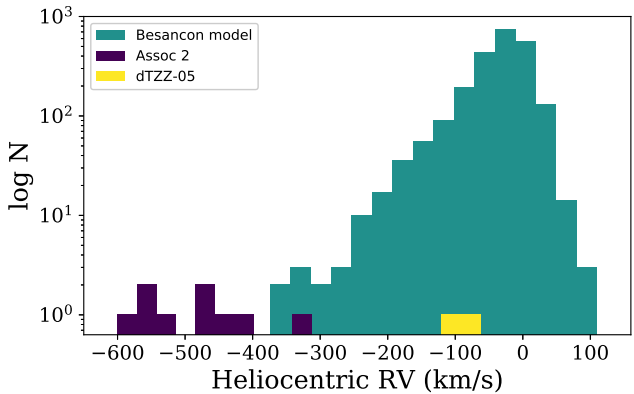}
\caption{A distribution of heliocentric radial velocities.  The candidate GC dTZZ-05 observed in this paper is shown in yellow.  Typical Milky Way field stars in the direction of dTZZ-05 from the Besan\c{c}on model of the Galaxy are shown in cyan.  M31 GCs that have been linked with ``Association 2'' are shown in purple \citep{Veljanoski2014,Mackey2019}.}\label{fig:dTZZ-05}
\end{center}
\end{figure}

Another cluster that is a clear outlier in Figure \ref{fig:ColorTest} is dTZZ-05.  The CaT-based [Fe/H] ratio here is higher than that predicted by its $(V-I)_0$ colour.  The GC dTZZ-05 was discovered fairly recently \citep{dTZZ}, and was not included in the kinematic analysis of \citet{Veljanoski2014}, meaning that there is no way to confirm whether the correct object was observed.  However, Figure \ref{fig:dTZZ-05} shows a histogram of the heliocentric radial velocity for dTZZ-05 from this work, along with Besan\c{c}on models of the Galaxy \citep{BesanconREF}\footnote{\url{http://model.obs-besancon.fr/}} in the direction of dTZZ-05.  The figure also shows the heliocentric radial velocities for the GCs that are classified as being in ``Association 2'' \citep{Veljanoski2014}, which \citet{Mackey2019} tentatively linked to dTZZ-05.  Figure \ref{fig:dTZZ-05} demonstrates that the object observed in this paper is clearly not a member of Association 2, but does have a radial velocity consistent with membership in the Milky Way.  This suggests either that dTZZ-05 is not a GC or that the wrong object was observed in this paper.  If the object that was observed in this paper is a foreground Milky Way star, that would explain its high [Fe/H] ratio.  Because of this discrepancy, dTZZ-05 is not included in subsequent discussions in this paper.

\section{Discussion}\label{sec:Discussion}
\begin{figure*}
\begin{center}
\centering
\hspace*{-0.15in}
\includegraphics[scale=0.6,trim=0.7in 0.0in 0.95in 0.in,clip]{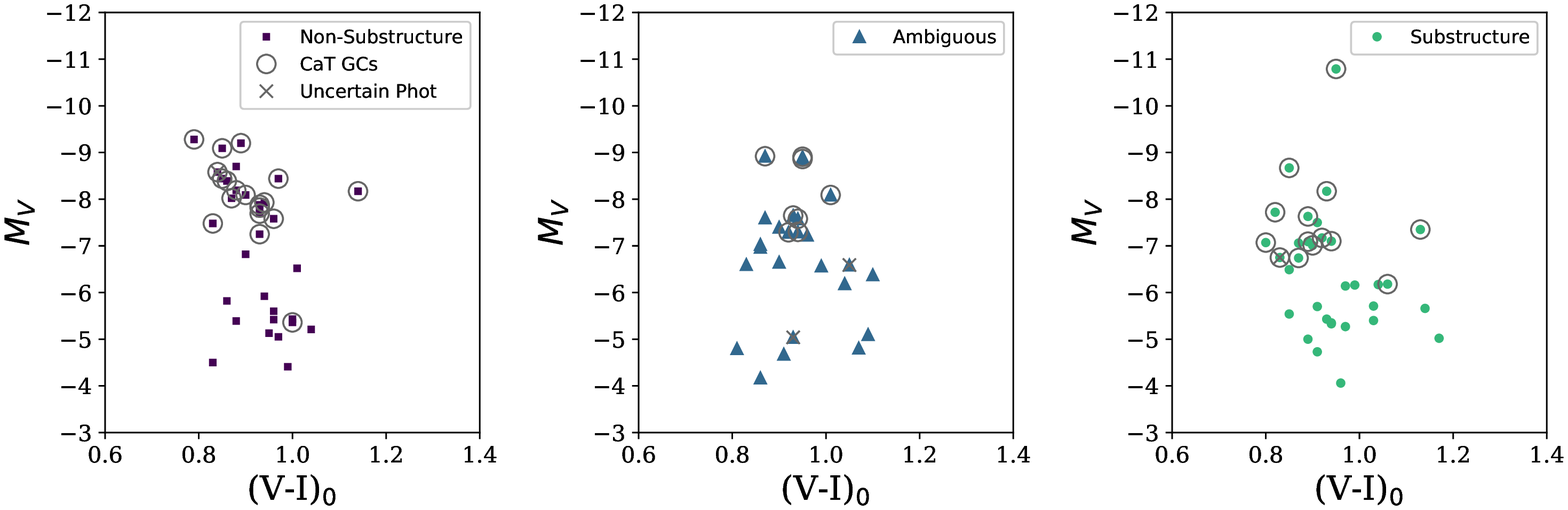}
\caption{\new{IL color-magnitude diagrams for outer halo M31 GCs, based on the absolute magnitudes and de-reddened colours from \citet{Mackey2019}.  GCs with uncertain photometry are indicated.  The GCs are divided into GCs associated with substructure (right), those that are not associated with substructure (left), and those that are ambiguous (middle).  The GC targets with CaT spectroscopy (this paper, \citealt{SakWall2016}, and \citealt{Sakari2021}) are circled.}}\label{fig:ILCMD}
\end{center}
\end{figure*}

Section \ref{sec:FeH} presented [Fe/H] ratios for these M31 GCs, along with estimates of uncertainties and comparisons with literature values.  \new{The CaT observations from this paper, \citet{SakWall2016}, and \citet{Sakari2021} yield metallicities for 58 of the 92 outer halo GC candidates \citep{Mackey2019}.  Figure \ref{fig:ILCMD} shows an IL colour-magnitude diagram of all outer halo GCs, demonstrating that these spectroscopic observations have larged probed the brightest outer halo GCs.  However, the spectroscopic observations have covered the full colour range for GCs associated with substructure, those with no associations with substructure, and those that are ambiguous.  Although this spectroscopic sample excludes fainter, more diffuse GCs (including the extended clusters; e.g., \citealt{Huxor2005}), the magnitudes of these GCs are similar to those that have been spectroscopically observed in the inner halo.}  Below the [Fe/H] ratios of these GCs are discussed in terms of what they reveals about M31's outer halo as a whole (Section \ref{subsec:M31OH}) and what they reveal about individual streams (Section \ref{subsec:Streams}).

\subsection{General Properties of M31's Outer Halo}\label{subsec:M31OH}
\begin{figure*}
\begin{center}
\centering
\hspace*{-0.15in}
\subfigure[]{\includegraphics[scale=0.53,trim=0in -0.1in 0.45in 0.2in,clip]{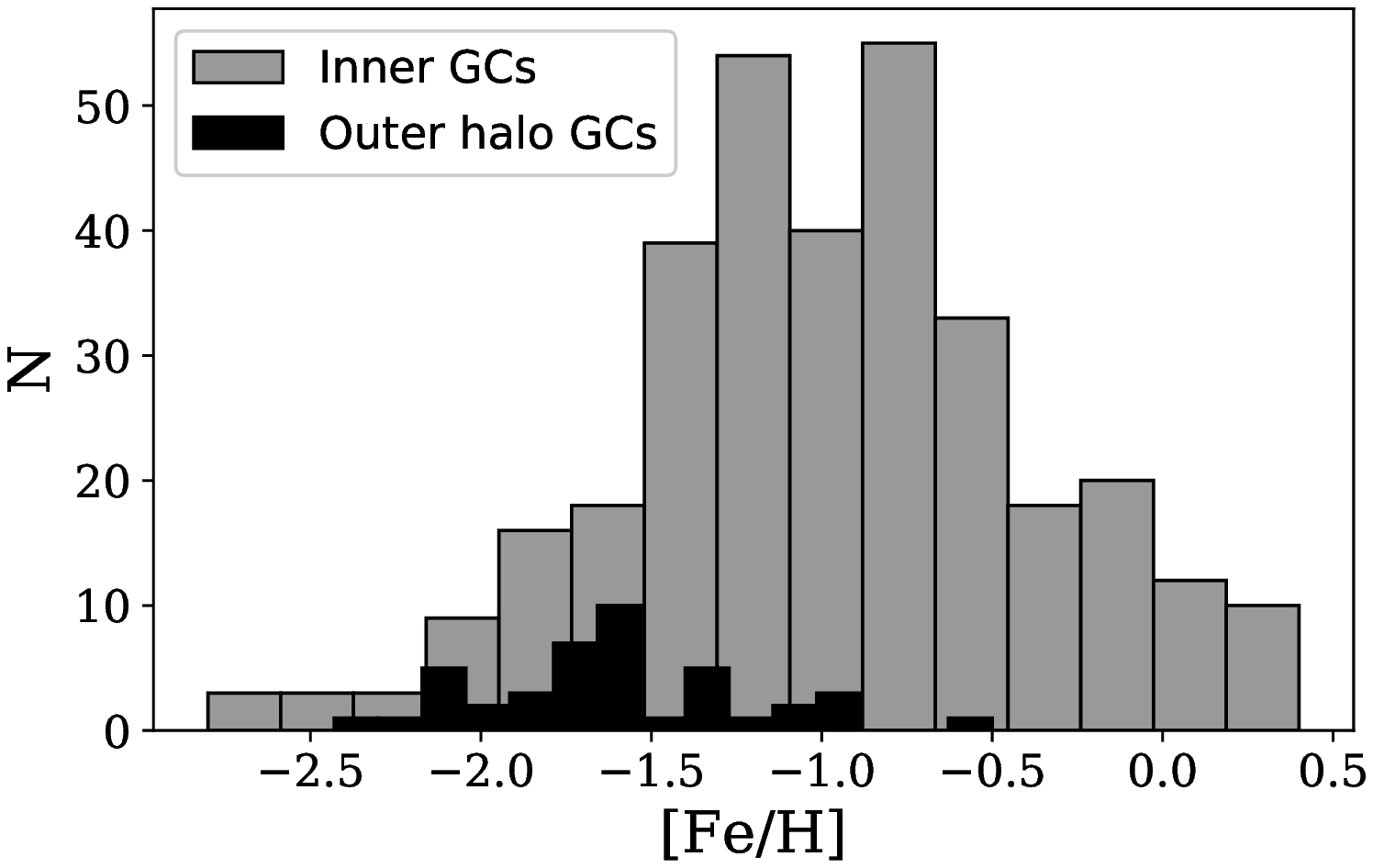}\label{fig:FeHDist}}
\subfigure[]{\includegraphics[scale=0.53,trim=0in -0.1in 0.45in 0.2in,clip]{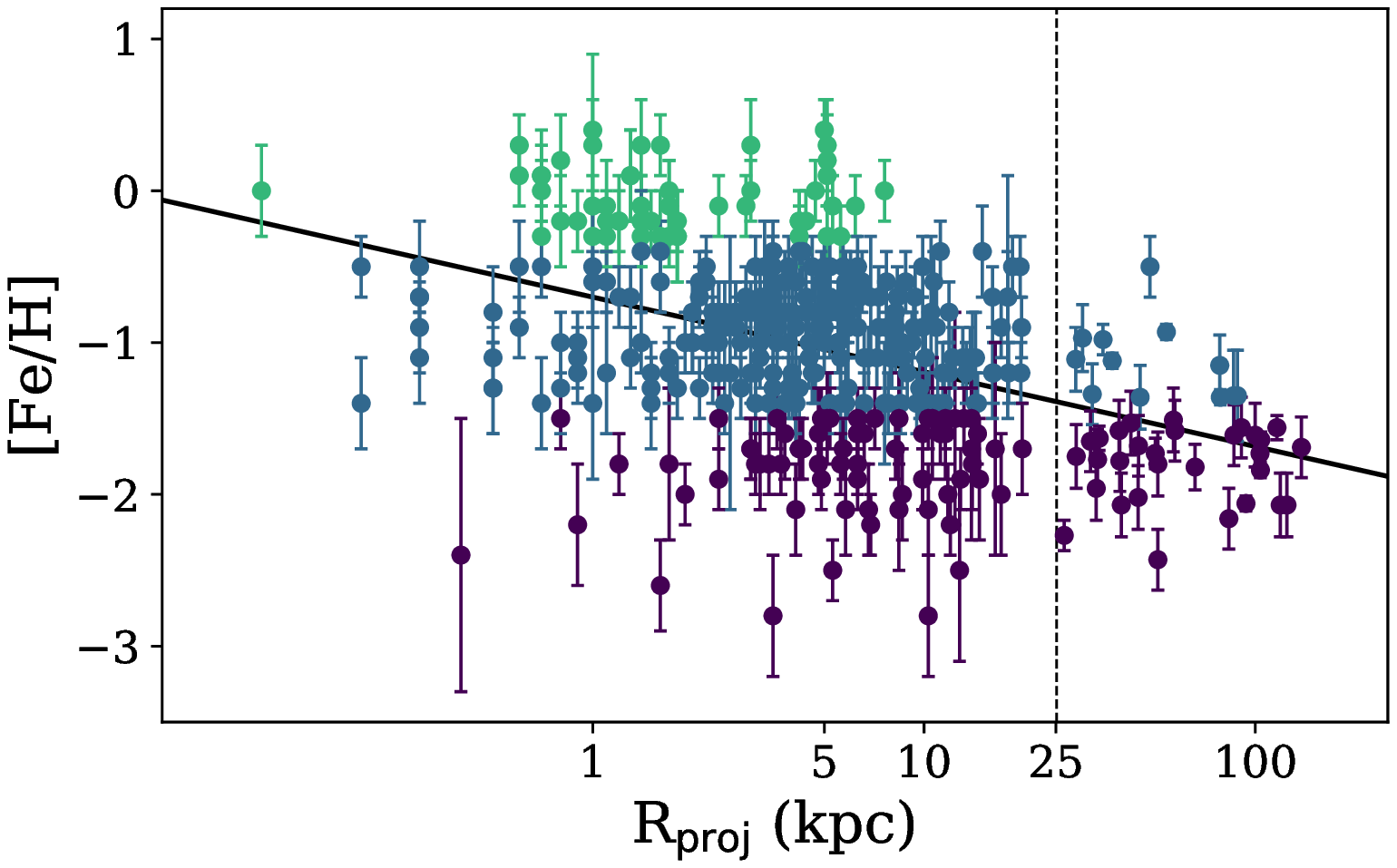}\label{fig:FeHgrad}}
\caption{Comparisons between old GCs in M31's outer ($R_{\rm{proj}} > 25$ kpc) and inner halo.  {\it Left: } Distributions of [Fe/H] ratios.  The inner halo sample is from \citet{Caldwell2011}, while the outer halo sample is from \citet{Colucci2014}, \citet{Sakari2015,Sakari2021}, and this paper (duplicate observations are only included once).  {\it Right: } GC [Fe/H] as a function of its projected distance from the centre of M31.  Again, the inner halo GC data is from \citet{Caldwell2011}, while the outer halo sample is from \citet{Colucci2014}, \citet{Sakari2015,Sakari2021}, and this paper.  Projected distances of outer halo GCs are from \citet{Mackey2019}.  The GCs are also colour-coded by [Fe/H] according to the metallicity groups from \citet{Caldwell2016}, where green are metal-rich GCs ($[\rm{Fe/H}]>-0.4$), blue are moderately metal-poor GCs ($-1.5 < [\rm{Fe/H}]<-0.4$), and purple are metal-poor GCs ($[\rm{Fe/H}]<-1.5$).  The solid line shows the metallicity gradient for all GCs, though note that there are not significant gradients within each [Fe/H] subpopulation.}
\label{fig:Comp}
\end{center}
\end{figure*}

Figure \ref{fig:FeHDist} shows the distribution of [Fe/H] ratios for inner and outer halo M31 GCs.  There are clearly fewer GCs in the outer halo, compared with the inner halo.  The outer halo is also more metal-poor on average than the inner halo: \new{the inner halo GCs have a mean $[\rm{Fe/H}]=-0.97\pm0.60$, compared to $[\rm{Fe/H}]=-1.61\pm0.40$ for the outer halo.} However, there are several fairly metal-rich GCs present in the outer halo, reaching up to $[\rm{Fe/H}]\sim -0.5$.  Previous analyses have shown that, unlike the Milky Way, M31's GC system is not clearly bimodal in [Fe/H], but instead seems to have multiple metallicity components \citep{Caldwell2011}.  There are too few GCs in the outer halo for a reliable Gaussian fit, however, the [Fe/H] spread is quite large given the small number of clusters. \new{Again, the presence of metal-rich GCs in the outer halo suggests accretion from a fairly massive dwarf galaxy (e.g., \citealt{Hughes2019}).}

Figure \ref{fig:FeHgrad} shows how GC metallicities change with projected distance from the centre of M31.  The dotted line shows the separation between the inner and outer halo at 25 kpc.  Previous studies have found metallicity gradients in M31 field stars (e.g., \citealt{Gilbert2014}).  Figure \ref{fig:FeHgrad} shows that there is an overall gradient in the M31 GCs (shown by the solid line).  However, the figure also shows the GCs colour-coded according to the metallicity groups of \citet{Caldwell2016}; there is no significant gradient in these individual sub-groups.  Instead, this figure demonstrates the changing populations of GCs with increasing distance from the center of M31.  The most metal-rich GCs in the outer halo (e.g., PA-37), which are conspicuous in this plot, are likely to have been accreted from dwarf satellites (see Section \ref{subsec:Streams}).

\begin{figure*}
\begin{center}
\centering
\hspace*{-0.15in}
\includegraphics[scale=0.6,trim=0.7in 0.0in 0.95in 0.in,clip]{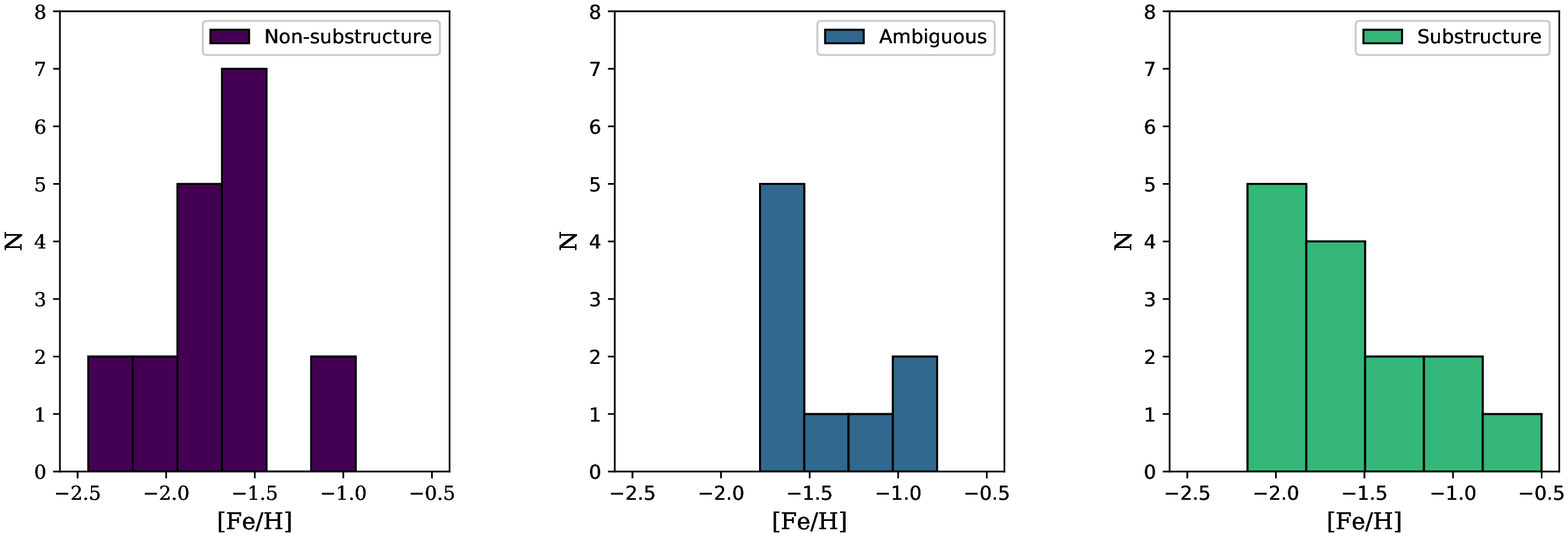}
\caption{\new{Distributions of GC metallicities in the non-substructured, ambiguous, and substructured components (according to the classifications from \citealt{Mackey2019}).  The [Fe/H] ratios are from this paper, \citet{Colucci2014}, and \citet{Sakari2015,Sakari2021}.}}\label{fig:FeH_Dist}
\end{center}
\end{figure*}

\subsection{Associations with Specific Streams}\label{subsec:Streams}
Several of the GCs have been linked to specific streams or substructure based on kinematics \citep{Veljanoski2014} or positions along stellar overdensities \citep{Mackey2019}, as listed in Table \ref{table:Targets}.  \new{Figure \ref{fig:FeH_Dist} shows the distribution of GC metallicities for GCs associated with substructure, those associated with the ``smooth'' (non-substructured) component, and those that have ambiguous associations with substructure \citep{Mackey2019}.  Simulations from \citet{Hughes2019} predict that the non-substructured GCs should be more metal-poor than those associated with substructure.  This is somewhat reflected in Figure \ref{fig:FeH_Dist}, where the most metal-rich outer halo GCs are in the substructure and ambiguous components (though note that there is one metal-rich GC, PA-17, that is categorized as a smooth, non-substructured GC).} Below, the properties of these GCs and their associated streams are discussed, based on the CaT-based metallicities.

\subsubsection{The Northwest Stream}\label{subsubsec:NWStream}
The Northwest (NW) Stream was evident in early PAndAS datasets \citep{McConnachie2009,Richardson2011}.  Although there are two components to the stream, K1 and K2, \citet{Preston2019} find that the two are likely unrelated.  K2 is the component with associated GCs, so K1 will not be discussed further here.  Simulations suggest that the progenitor of the K2 component was likely massive, with stellar mass estimates ranging from $>2.2\times10^6$~M$_{\sun}$ \citep{Kirihara2017} to $8.5\times10^6$~M$_{\sun}$ \citep{McConnachie2018}.  For reference, the stellar mass of the Fornax dwarf spheroidal galaxy, a satellite of the Milky Way, is $M\sim 20\times10^6$~M$_{\sun}$ \citep{McConnachie2012}.  \citet{Ibata2014} find that that the stream itself is generally moderately metal-poor, with $-1.7 \la [\rm{Fe/H}] \la -1.1$.

There are 6-7 GCs associated with the K2 NW Stream component.  This high number of GCs leads to a very high specific frequency for the NW Stream, $S_N = 70-85$, considerably higher than the Fornax dwarf spheroidal \citep{Mackey2019}.  The three NW Stream GCs analyzed here, PA-04, PA-09, and PA-11, are all relatively metal-poor, with $[\rm{Fe/H}]<-1.5$.  These CaT-based metallicities are consistent with resolved photometry by \citet{Komiyama2018}, who found PA-10, -11, -12, and -13 to be metal-poor, similar to the surrounding stream; however, they also found that PA-11 and -13 may be even more metal-poor than PA-10 and -12.  With a CaT-based metallicity of $[\rm{Fe/H}]=-2.16$, PA-11 is indeed more metal-poor.  PA-04, the brightest GC of the NW Stream, is similarly metal-poor, at $[\rm{Fe/H}]=-2.07$; these relative metallicities are consistent with their IL colours.  PA-09 is found to be slightly more metal-rich, at $[\rm{Fe/H}]=-1.56$.\footnote{Note that photometric measurements of PA-09 are uncertain because of the presence of a nearby bright star \citep{Komiyama2018}.  The bright star was avoided for the long slit spectroscopy in this paper.}

The K2 component of the NW Stream therefore appears to have at least five metal-poor GCs, with one that is as metal-rich as $[\rm{Fe/H}]=-1.56$.  In this sense, the K2 component may be similar to the Fornax dwarf spheroidal, which has four GCs with $[\rm{Fe/H}]\la -2$ \citep{Letarte2006} and one at $[\rm{Fe/H}]=-1.4$ \citep{Hendricks2016}.  \citet{Hendricks2016} found that the metal-rich GC had lower [$\alpha$/Fe] than the other GCs, based on Si, Ca, and Ti abundances; the GCs therefore trace the chemical evolution of the Fornax field stars.  If the NW Stream progenitor was similarly massive, PA-09 could also have lower [$\alpha$/Fe]; follow-up observations are necessary to confirm this.

\subsubsection{The Southwest Cloud}\label{subsubsec:SWCloud}
The Southwest (SW) Cloud was evident in early PAndAS data \citep{McConnachie2009} and has since been analyzed in detail by \citet{Bate2014} and \citet{McMonigal2016}.  \citet{Bate2014} found the stream to have a mean $[\rm{Fe/H}]~\sim~-1.3$, with a current mass of $7.1\times10^6$~M$_{\sun}$; however, they argue that the original progenitor was likely much more massive, around $20\times10^6$~M$_{\sun}$, i.e., about the mass of the Fornax dwarf spheroidal.

As with the NW Stream, \citet{Mackey2019} find the SW Cloud to have a very high specific frequency of $S_N = 90-150$, based on its current mass.  There are three GCs projected along the stream, along with two more that lie on an extension of the stream.  Of these five GCs, all are moderately metal-poor.  From resolved photometry, \citet{Mackey2013} find PA-7 and -8 to have $[\rm{Fe/H}]~\sim~-1.3$, with signs of younger ages.  From high-resolution IL spectroscopy, \citet{Sakari2015} found one of the GCs on the extension, H10, to also have $[\rm{Fe/H}]~\sim~-1.3$, with elevated [Ca/Fe] that indicates it originated in a fairly massive progenitor galaxy.  This paper adds a metallicity for the third GC along the stream, PA-14, which is found to have a CaT-based metallicity of $[\rm{Fe/H}]~=~-1.6$.  Although slightly more metal-poor than the other SW Cloud GCs, its metallicity is consistent with the population in the SW Cloud stream itself \citep{Ibata2014}.

It is intriguing that none of the GCs associated with the SW Cloud are very metal-poor, especially given the cloud's high specific frequency. \citet{McConnachie2018} argue that the SW Cloud could be a ``shell'' of material from a larger merger; if this is true, the SW Cloud GCs may have been accreted from a more massive galaxy (which would be consistent with H10's elevated [Ca/Fe]; \citealt{Sakari2015}).  Additional follow-up of these GCs would help characterize the nature of the SW Cloud.

\subsubsection{Association 2}\label{subsubsec:Assoc2}
Association 2 is a grouping of GCs that was first identified by \citet{Mackey2010b}.  Although there is no obvious association with a stellar stream, the GCs appear to be clustered together in projection.  \citet{Veljanoski2014} found that there are two kinematic subgroups in Association~2, each of which is also clustered in position.  Although G1 and G002 have projected locations nearby, their radial velocities are distinct enough to indicate that they are likely not part of either subgroup in Association~2.\footnote{Note that dTZZ-05 has been tentatively linked to Association~2 \citep{Mackey2019}.  The radial velocity in this paper shows that the object observed in this paper is not associated with Association~2 (Figure \ref{fig:dTZZ-05}); however, the observed object may be a foreground field star.}

This paper has analyzed two GCs from one of the Association~2 kinematic subclumps: H7 and PA-22.  These are the two brightest GCs in their subgroup.  \citet{Mackey2019} found some evidence that both of these GCs may lie along unnamed metal-poor substructure.  This paper finds both GCs to be moderately metal-poor, at $[\rm{Fe/H}]~=~-1.34$ and $-1.11$, respectively.  The H7 metallicity is roughly in agreement with the previous value from \citet{Wang2021}, though it is much more metal-rich than found by \citet{Wang2019}; PA-22's CaT-based metallicity is roughly consistent with \citet{Wang2019}. (Note that Wang et al. find PA-22 to be slightly more metal-rich, with an age $\sim5$ Gyr).  These GCs are fairly metal-rich for outer halo GCs.  \citet{Veljanoski2014} noted that this kinematic clump has similar velocities as the outer disk (e.g., \citealt{Ibata2005}), indicating that the GCs may be associated with the disk.  Future high-resolution spectroscopic follow-up of these GCs will better help constrain the properties of their birth environment.

\subsubsection{The Giant Stellar Stream}\label{subsubsec:GSS}
The Giant Stellar Stream (GSS) is the most notable substructure in the outer halo.  The GSS was first discovered by \citet{Ibata2001}, and subsequent PAndAS imaging revealed that the metal-rich outer halo is dominated by the GSS \citep{Ibata2014}.  \citet{McConnachie2018} find that the GSS accounts for 93\% of the mass of the ``named'' (i.e., significant and bright) substructure in the outer halo.  From photometry, \citet{Conn2016} determined the GSS to be metal-rich, with the bulk of the stars having $-0.7\la[\rm{Fe/H}]\la-0.2$.    \citet{Escala2021} further find GSS stars to have elevated [$\alpha$/Fe] ratios, suggesting that they originated in a massive galaxy.

Models of the progenitor and its orbit show that whatever created the GSS was massive: \citet{Fardal2006} estimate a stellar mass of $M_{*} \sim 10^9\;M_{\sun}$; \citet{Kirihara2017a} find that the progenitor was a spiral galaxy, with a similar mass as the Large Magellanic Cloud; and \citet{Hammer2018} argue that the progenitor was as massive as 25\% of M31.  Such a significant merger was likely responsible for other features in M31 as well.  The models of \citet{Fardal2006,Fardal2013} indicate that the progenitor orbited several times, leading to multiple wraps around M31.  Several other features have been attributed to the GSS, including the eastern and western shelves (e.g., \citealt{Ferguson2002,Ferguson2005}) and the eastern extent near the GSS \citep{Preston2021}.

Despite its origins as a massive satellite galaxy, the GSS has a conspicuous lack of GCs in the outer halo.  \citet{Mackey2019} identify three GCs that lie along the GSS in projection: PA-37, H19, and H22, all of which are analyzed in this paper.  Based on their radial velocities and positions (specifically, the lack of background metal-rich stars in the immediate vicinity), \citet{Mackey2019} conclude that H19 and H22 are likely not associated with the GSS, though PA-37 is possibly associated with the stream (also see \citealt{Preston2021}).  The CaT analysis in this paper finds PA-37 to be metal-rich ($[\rm{Fe/H}] = -0.50$), consistent with the GSS stars from \citet{Conn2016}.  PA-37 is therefore the most metal-rich outer halo GC in this sample.  Such a high metallicity is unusual for an outer halo GC, further supporting an association with the GSS.  On the other hand, H19 and H22 are considerably more metal-poor; their radial velocities also agree with those from \citet{Veljanoski2014}, supporting the claim from \citet{Mackey2019} that they are not associated with the GSS.

\new{There is therefore only one outer halo GC that seems to be clearly associated with the GSS.}  For comparison, the LMC has at least 16 classical GCs (e.g., \citealt{Harris2013}).  One possibility is that the GSS progenitor did have more GCs, but that they were stripped early on and are currently projected onto previous wraps around M31 or into the inner halo (e.g., \citealt{Mackey2019}). \new{Simulations suggest that massive satellites will tend to deposit their GCs in the inner regions \citep{Pfeffer2020}.}  If the GSS GCs are currently located elsewhere in the outer halo, this could explain the high specific frequency of the outer halo \citep{Mackey2019}.  There are also other metal-rich GCs that are currently projected into the outer halo.  For instance, \citet{Sakari2015} identified the outer halo GC PA-17 as moderately metal-rich ($[\rm{Fe/H}] \sim-0.9$) with low [Ca/Fe] that indicated it formed in a dwarf galaxy.  PA-36 is found to be similarly metal-rich (see Section \ref{subsubsec:Other}).  If PA-17 or PA-36 came from the GSS progenitor, they must have been accreted on an earlier orbit around M31.  Further dynamical work and spectroscopic observations will be essential for assessing the nature of the GSS and its GC(s).

\subsubsection{Streams C and D}\label{subsubsec:StreamsCD}
Streams C and D are two separate streams which intersect in the north and merge with the GSS in the south.  Various parts of the stream were first identified by \citet{Ibata2007}, \citet{McConnachie2009}, and \citet{Richardson2011}.  Based on the initial observations, \citet{Ibata2007} argued that the progenitors were more massive than the Fornax dwarf spheroidal. The two streams appear to have different stellar populations, where Stream D seems to be more metal-poor than C \citep{Ibata2014}.  \citet{Chapman2008} found that Stream C is composed of two separate components with distinct radial velocities: the Cp component is more mteal-poor ($[\rm{Fe/H}] \sim-1.3$) than the Cr component ($[\rm{Fe/H}] \sim -0.7$).  From a handful of candidate Stream D stars, \citet{Chapman2008} find Stream D to be moderately metal-poor ($[\rm{Fe/H}] \sim-1.4$).

There are many GCs projected onto Streams C and D.  \citet{Mackey2019} find three GCs projected onto Stream C, three onto D, and another nine onto the overlapping C/D region. This paper includes five of the nine GCs in this overlapping region.  \citet{Veljanoski2014} further found that the C/D GCs could be kinematically separated into two subgroups.  Three of the GCs in this paper (H24, PA-41, and PA-46) are in the first of the kinematic subgroups, while another two (B517 and PA-44) are in the second subgroup.

For the first subgroup, this analysis finds similar radial velocities as \citet{Veljanoski2014} for H24 and PA-46, though the velocity for PA-41 is about 40 km/s lower than the \citet{Veljanoski2014} value.  All three of the GCs in this subgroup have radial velocities that are offset from the Stream C and D velocities in \citet{Chapman2008}, though those velocities are from fields further to the south.  Chapman et al. do find a possible velocity gradient in the Cp stream, which could match these C/D GCs.  All of the GCs in this first subgroup are also metal-poor: PA-41 and PA-46 have $[\rm{Fe/H}]\sim-2$, while H24 is slightly more metal-rich, at $[\rm{Fe/H}]=-1.58$.  H24's slightly higher [Fe/H] is consistent with its redder IL $(\rm{V-I})_0$ colour and agrees with the metallicity from \citet{Wang2021}.  The similar velocities and metallicities of these GCs suggests that they originated in a galaxy that was massive enough for ongoing star formation and formation of multiple GCs.

The second subgroup identified by \citet{Veljanoski2014} has a larger velocity spread, so it is not as clear if the GCs shared a common birth environment.  The [Fe/H] ratios for the two targets in this subgroup are very different.  PA-44 is found to be more metal-poor, with $[\rm{Fe/H}]\sim-2$, while B517 is more metal-rich, at $[\rm{Fe/H}]=-1.36$.  Again, the C/D overlap region is further away from the C and D fields analyzed by \citet{Chapman2008}, making it difficult to assess membership with either stream based on radial velocities.  However, B517's radial velocity is very discrepant from the Stream Cr and D velocities; \citet{Chapman2008} also found no evidence for a velocity gradient in the those streams.  Although B517's velocity is similar to the Stream Cp velocities from \citet{Chapman2008}, it is inconsistent with the observed velocity gradient.  PA~44's radial velocity is more similar to Stream D, but again, any comparisons with either stream are tenuous.  Ultimately, more work is necessary to probe the nature of Streams C and D and their associations with the GCs.

\subsubsection{Other Substructure}\label{subsubsec:Other}
There are a number of GCs that \citet{Mackey2019} classify as ``substructure'' GCs despite no obvious association with a named stream.  These GCs have a significant background density, indicating that they may lie along undetected, metal-poor streams.  Two of the GCs in this study, PA-36 and G339, fall into this category.  G339 is found to be fairly metal-poor, at $[\rm{Fe/H}]=-1.75$, while PA-36 is more metal-rich, at $[\rm{Fe/H}]=-0.97$.  PA-36 is quite metal-rich for an outer halo GC.  Figure \ref{fig:PA36} compares the PA-36 spectrum to other M31 GCs, demonstrating that its IL CaT lines are consistent with high [Fe/H].  However, PA-36's bluer IL $(\rm{V-I})_0$ colour hints at a more metal-poor GC (Figure \ref{fig:ColorTest}). It is worth noting that PA-36 has a similar IL $(\rm{V-I})_0$ colour and CaT-based [Fe/H] as the massive GC G1 \citep{Sakari2021}.  G1 is known to have a significant spread in He and/or Fe \citep{Nardiello2019}; it is possible that PA-36 may also have similar abundance spreads, even though it is less massive than G1. \new{Alternatively, PA-36 may be younger than the other outer halo GCs.}

Given their proximity to the bottom of the K1 component of the NW Stream and And XXVII, it is worth considering whether the GCs could be associated with either.  However, \citet{Preston2019} find very different radial velocities for the stream and And XXVII.  Although the stream does have a velocity gradient, the direction of the gradient is inconsistent with the two GCs (see Figure 8 in \citealt{Preston2019}).  Instead, it is possible that the two GCs were brought in by older accretion evens or older orbits around M31, such that the associated streams have dissolved somewhat and become harder to distinguish.

\subsubsection{GCs with no associated substructure}\label{subsubsec:Nosubs}
Thirteen of the GCs in this paper are classified as ``non-substructure'' by \citet{Mackey2019}.  This classification is due to a combination of the lower background stellar density surrounding the GCs and positions and radial velocities that are not obviously connected with known streams, GCs, or satellite galaxies.  These GCs also seem to follow the radial distribution of the ``smooth,'' metal-poor halo \citep{Ibata2014,Mackey2019}.  \citet{Mackey2019b} showed that these non-substructure GCs have different rotational signatures from the substructure GCs, suggesting that they came from \new{ancient accretion events.}

Of the non-substructure GCs in this paper, most are found to be metal-poor.  PA-16 is in this group (though see the discussion in Section \ref{subsubsec:PA16}), as is the very metal-poor GC EXT~8.  The metallicities of most of these GCs are similar to the predicted metallicities for the smooth stellar halo \citep{Ibata2014}.

However, some non-substructure GCs are more metal-rich, unlike the non-substructure (``smooth'') outer halo stars, which have with $[\rm{Fe/H}]<-1.7$ \citep{Ibata2014}.  PA~52 is one such cluster; at $[\rm{Fe/H}]=-1.15$, it is significantly \new{more metal-rich} than the smooth halo population. PA-52 lies near the Eastern Cloud, along a possible extension of stars \citep{McMonigal2016} and has a radial velocity that is somewhat consistent with other Eastern Cloud GCs, hinting at a possible association. However, another metal-rich GC, PA-17 \citep{Sakari2015}, has also been classified as a non-substructured GC \citep{Mackey2019}; it does not have an obvious progenitor.  The massive, moderately metal-rich GC G1 also does not have an obvious host galaxy (see, e.g., \citealt{Sakari2021}).  It remains unclear if these metal-rich GCs can be associated with the ancient accretion event that brought in the more metal-poor GCs and field stars.

\section{Conclusions}\label{sec:Conclusion}
This paper has provided [Fe/H] ratios for thirty GCs in the outer halo of M31.  These metallicities were determined from an empirical relationship between Voigt profile fits to the second of the near-infrared CaT lines and IL [Fe/H], based on a previously-analyzed sample of M31 GCs \citep{SakWall2016}.  Although previous analyses find that the CaT is a better tracer of a GC's [Ca/H] ratio, this paper finds no significant difference between using [Ca/H] or [Fe/H]; for ease of interpretation, [Fe/H] is used here.  Several of these [Fe/H] ratios are found to disagree with results from the literature, which may indicate issues with GC ages, undetected features of the underlying stellar populations, and/or problems with the CaT-[Fe/H] relationship at very low metallicities.  However, the majority of these GCs have CaT-based [Fe/H] ratios that are consistent with IL colours and literature results.

These [Fe/H] ratios lead to several key findings about M31's assembly history, as summarized below.
\begin{itemize}
    \item Many of the outer halo GCs are metal-poor, consistent with the ``smooth'' (non-substructured) population of outer halo field stars.  This is particularly true for the GCs that are not associated with any substructure, which (with one exception) all have $[\rm{Fe/H}]<-1.5$.
    \item Several named substructures (the NW Stream and the SW Cloud) are shown to have GCs with metallicity spreads, indicating that they experienced ongoing, extended star formation.  Based on the metallicities, the NW Stream's GC population bears some resemblance to the Fornax dwarf spheroidal.
    \item The one GC that is likely connected to the GSS, PA-37, is found to be very metal-rich for an outer halo GC.  At $[\rm{Fe/H}] = -0.50$, it is the most metal-rich GC in the outer halo.  It is unclear if other GCs are associated with the GSS.
    \item There are several moderately metal-rich GCs (e.g., PA-36 and PA-52) that don't have an obvious host galaxy.  However, GCs with $[\rm{Fe/H}] \sim -1$ are unusual in the outer halo, suggesting that they were likely accreted from satellite galaxies.
    \item Overlapping streams and groups of GCs (e.g., those in Streams C/D and Association 2) are difficult to interpret.  The distinct kinematic groups imply that the subpopulations of GCs originated in different environments.  The [Fe/H] ratios further complicate these analyses, showing that in some cases the host galaxies had to have been massive enough for extended star formation, creating GCs with metallicity spreads.
    \item Several GCs, including PA-16 and PA-36, also have intriguing offsets between their CaT-based [Fe/H] ratios, IL colours and literature results.  These offsets could be the result of interesting underlying stellar populations or younger ages.  Additional observations are needed to continue to investigate these GCs.
\end{itemize}

The CaT-based [Fe/H] ratios have provided insight into the nature of M31's outer halo GCs and the assembly history of the outer halo as a whole.  Additional follow-up will help further constrain the properties of the outer halo GCs and their birth environments.

\section*{Data Availability}
Spectra of GCs will be made available upon request.

\section*{Acknowledgments}
CMS dedicates this paper to George Wallerstein in gratitude for the tremendous support he provided in her postdoc years and for mentoring her in the usage of puns (which was surely one of his clowning achievements).

CMS also thanks the anonymous reviewer for suggestions that have improved the manuscript.

The authors also thank the observing specialists at Apache Point Observatory and McDonald Observatory for their assistance with these observations.

GW acknowledges funding from the Kenilworth Fund of the New York
Community Trust.
This research has made use of the SIMBAD database, operated at CDS,
Strasbourg, France.

\footnotesize{

}


\end{document}